\documentclass[12pt]{article}
\usepackage{epsfig}
\usepackage{graphics,graphicx}
\usepackage{amssymb}
\usepackage{amsmath}
\renewcommand{\baselinestretch}{1.5}

\newcommand{\bm}[1]{\mbox{\boldmath$ #1 $\unboldmath}}
\def\qed{\hfill$\diamondsuit$}
\begin{document}
\begin{center}
{\Large\bf Analysis of Computer Experiments with Functional Response}\\
Ying Hung, V. Roshan Joseph$^{\dag}$, and Shreyes N. Melkote$^{\ddag}$\\
\bigskip
Department of Statistics and Biostatistics,\\
Rutgers, the State University of New Jersey, Piscataway, NJ\\
\bigskip
$^{\dag}$School of Industrial and Systems Engineering,\\
Georgia Institute of Technology, Atlanta, GA 30332\\
\bigskip
$^{\ddag}$School of Mechanical Engineering,\\
Georgia Institute of Technology, Atlanta, GA 30332\\
\bigskip
\end{center}
\renewcommand{\baselinestretch}{1.5}
\begin{abstract}

This paper is motivated by a computer experiment conducted for optimizing residual stresses in the machining of metals. Although kriging is widely used in the analysis of computer experiments, it cannot be easily applied to model the residual stresses because they are obtained as a profile. The high dimensionality caused by this functional response introduces severe computational challenges in kriging. It is well known that if the functional data are observed on a regular grid, the computations can be simplified using an application of Kronecker products. However, the case of irregular grid is quite complex. In this paper, we develop a Gibbs sampling-based expectation maximization algorithm, which converts the irregularly spaced data into a regular grid so that the Kronecker product-based approach can be employed for efficiently fitting a kriging model to the functional data.

\end{abstract}
\newpage
\begin{center}
{\large\bf 1. INTRODUCTION}
\end{center}

Computer experiments (Santner et al., 2003; Fang et al., 2006)
refer to those experiments that are performed using computers with the help of
physical models and numerical methods, such as finite element analysis.
Many computer experiment
responses are collected in a functional form. That is, for each
setting of the experiment, responses are collected over an interval of some index,
such as space or time. The study of functional data is important because it
can help us understand how the factors affect the shape of the
resulting curve, which can have a bearing on the performance of the object
or system under investigation.

\begin{figure}
\centering\resizebox{300pt}{250pt}
{\includegraphics{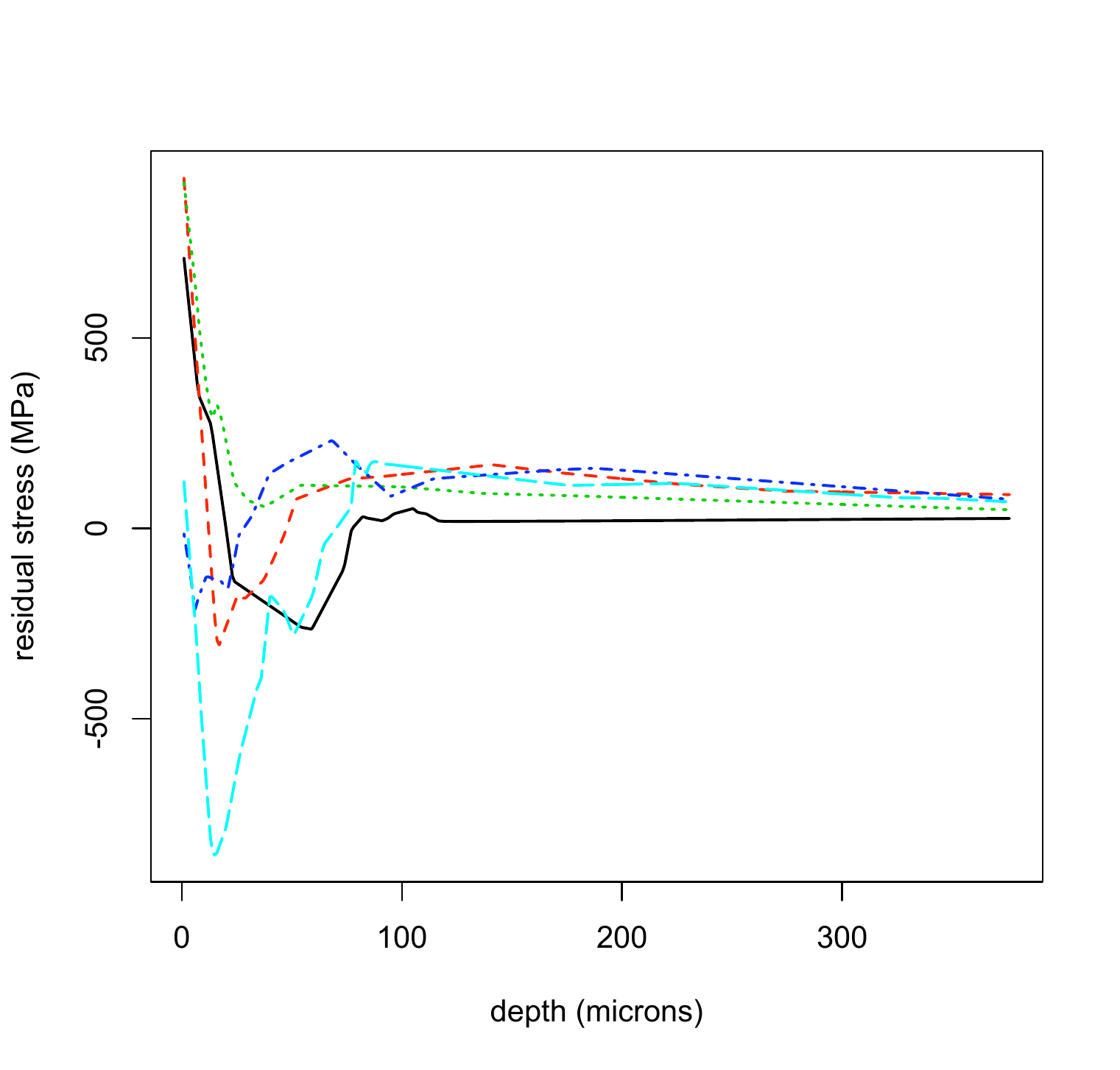}} \caption{Five randomly sampled residual stress
profiles vary with the settings.}\label{RESPROFILE}
\end{figure}

This paper is motivated by a
computer experiment in the machining of metals  using a hard turning process.
The main objective of the experiment is to study the
residual stresses generated in the machined surface because they are known to influence the fatigue life
and are also associated with distortion in
machined parts. Nine
machining process variables are considered in this simulation experiment and for
each setting, the residual stress profiles over the depth at three different locations
in the machined part are generated. Specifically, each residual
stress profile measured in the tool feed rate direction contains 376
residual stress values output by the simulation code over a depth of
376 microns measured from the part surface (i.e. one residual stress value per micron). Figure
\ref{RESPROFILE} illustrates five sample residual stress profiles obtained under
different machining process settings. We can see that the residual stress
profiles vary with the settings. Therefore, it is
important to predict the effect of the process variables on the
residual stress profile so that the machining process can be
optimized to enhance the fatigue life of machined components.

The literature on modeling computer experiments with functional
responses is scant as most of the modeling techniques focus on
single or multiple outputs (Conti et al. 2009, Conti and O'Hagan 2010). Kriging (Sacks et al. 1989) is the most popular
technique for modeling these data due to its interpolating property, which is desirable in deterministic computer experiments.
However, kriging is not used for analyzing functional data
because of the computational problems caused by the high dimensionality of functional responses, especially when they are
collected with an intensive sampling rate. Techniques such as wavelet decomposition (Bayarri et al. 2007),
principal component analysis (PCA) (Ramsay and Silverman 2005, Higdon et al. 2007), functional linear models (Fang et al. 2006), and knot-based Gaussian process models (Banerjee et al. 2008) have been used instead. Although they provide simple and fast solutions, these models cannot interpolate the data. This creates a mismatch in the analysis methods for single and functional outputs. Thus, the main objective of this work is to improve the computational efficiency of kriging in the analysis of functional data so that the same method can be applied irrespective of the type of the response.

A naive extension of kriging to functional
response is to include the functional argument as an
additional input to the model (Kennedy and O'Hagan 2001, Liu and West 2009). For example, the machining experiments are conducted
with nine process variables and a spatial variable that
indicates the locations of the residual stress measurements. The kriging model can be fitted by incorporating
depth as the 11th variable. Although this method is simple, it
suffers from the computational difficulties. This is because the maximum likelihood estimation of the correlation parameters in the kriging model involves the inversion and determinant calculation of a
correlation matrix, whose dimension increases with the total number
of observations $N=nm$, where $n$ is the run size of the
experiment, and $m$ is the number of observations (from the
functional space) in each run. These large matrix operations make
the estimation of the correlation parameters computationally intensive and numerically unstable (Joseph and Kang 2011). Take the 90-run
machining experiment as an example; the inversion and determinant
calculation of a $33840 \times 33840$ matrix ($n=90$, $m=376$, and
$N=33840$) is required at each iteration of the optimization algorithm used in the maximum likelihood estimation, which will make the optimization extremely time-consuming.

One approach to overcome the computational issues associated with the naive
kriging extension is to apply a Kronecker product formulation for constructing the correlation matrices (Williams et al. 2006, Rougier 2008, Liu et al. 2008, Bayarri et al. 2009).
However, this approach can only be used  when the functional responses are collected over a regular grid, which means that the outputs are observed at the same locations in the functional space for all runs.


Although not as common as the case of regular grid, observations on nonregular grid also occur sometimes in practice. For examples, in the study of transient rolling adhesion and deformation of leukocytes, the displacement profile of a cell is simulated using computational fluid dynamic models. These profiles are often truncated irregularly at various time points due to computational constraints at different kinetic parameter settings (Pappu and Bagchi 2008).
In the thermo-mechanical study of the friction drilling processes, the thrust force profiles are generated from finite element modeling over the distance the tool travels. The profiles are irregularly collected because the travel distances are different at different values of tool feed rate (Miller and Shih 2007). In the motor engine simulations reported in Liu et al. (2008), the acceleration profiles are truncated at different time points leading to functional data on an irregular grid. Such truncated profiles are also commonly seen in degradation studies and fatigue testing simulations because the profile data cannot be collected when the product fails.

In this work, we propose a general and efficient method to overcome the computational issue in analyzing functional response, especially with irregular grid. The paper is organized as follows. In Section 2, a brief review of the kriging model is given.
The new modeling procedure for functional response in computer experiments is developed in Section 3. The proposed method is illustrated using the machining experiment in Section 4. Summary and concluding
remarks are given in Section 5.

\begin{center}
{\large\bf 2. Kriging Preliminaries}
\end{center}

Suppose that the computer experiment is conducted using $p$ variables $\bm x=(x_1,\cdots,x_p)'$
and the functional responses are collected
over an index $t$. For each experimental setting
$\boldsymbol{x}_i \in \mathcal{R}^p$, $i=1,\cdots,n$, the  outputs
$\boldsymbol{y}_i=(y_{i1}, \cdots, y_{im_i})'$ are assumed to be a
vector which is observed over $t_{i1},\cdots, t_{i\,m_i}$ with ${\bf t}_i=(t_{i1}, \cdots, t_{i\,m_i})'$.
Note that the functional responses for each run can be collected differently in terms of location,
therefore, ${\bf t}_i$'s and $m_i$'s are not necessarily the same for every $i$. Then, the kriging model is given by
\begin{equation}\label{OK2}
y(\boldsymbol{x},t)=\boldsymbol{\upsilon}(\boldsymbol{x},t)'\boldsymbol{\mu}+Z(\boldsymbol{x},t),
\end{equation}
where $y(\boldsymbol{x},t)$ is the response at point $t$
and input setting $\boldsymbol{x}$,
$\boldsymbol{\upsilon}(\boldsymbol{x},t)'=(\upsilon_0(\boldsymbol{x},t),
\upsilon_1(\boldsymbol{x},t),\cdots,\upsilon_L(\boldsymbol{x},t))$ is a set of known functions (usually, $\upsilon_0(\boldsymbol{x},t)=1$), and $\bm \mu$ is a vector of unknown parameters. We assume that $Z(\boldsymbol{x},t)$ is a
Gaussian process with mean 0 and the covariance function
$cov\{y(\boldsymbol{x}_1,t_1),y(\boldsymbol{x}_2,t_2)\}=\sigma^2
r(\boldsymbol{x}_1-\boldsymbol{x}_2,t_1-t_2)$. Furthermore, it is common to assume  a
separable product correlation function:
$r(\boldsymbol{x}_1-\boldsymbol{x}_2,t_1-t_2)=
(\Pi_{i=1}^p r_i(x_{i1}-x_{i2}))r_T(t_1-t_2)$, where
$r_i(x_{i1}-x_{i2})$ is the correlation function for the $i$th
variable and $r_T(t_1-t_2)$ is the correlation function for
variable $t$. Unknown parameters associated with these correlation
functions are denoted by $\boldsymbol{\xi}$. Note that the
spatial-temporal model (Cressie 1993, Fang et al. 2006) is a special case of
this model where $\boldsymbol{x}$ represents space
and $t$ represents time.

Suppose the $N\times 1$ vector $\boldsymbol{y}=(\boldsymbol{y}'_1,\cdots,\boldsymbol{y}'_n)'$ 
is a collection of all the functional outputs, where $N=\sum_i^n m_i$. The experimental design is
denoted by the $N \times p$
matrix ${\bf X}=({\bf 1}'_{m_1} \otimes \boldsymbol{x}_1,\cdots, {\bf 1}'_{m_n} \otimes \boldsymbol{x}_n)'=
({\bf X}_1,\cdots,{\bf X}_N)'$, where $\otimes$ denotes
the Kronecker product operator, ${\bf 1}_{m_i}$ is a column of 1's having length $m_i$, and ${\bf T}=({\bf t}_1',\cdots, {\bf t}_n')'=(t^*_1,,\cdots,t^*_N)'$ is
the corresponding $N\times 1$ vector on the functional space.
Based on model (\ref{OK2}), the universal kriging predictor is
given by
\begin{equation}\label{KRIGPRED}
\hat{y}(\boldsymbol{x},t)=
\boldsymbol{\upsilon}(\boldsymbol{x},t)'\hat{\boldsymbol{\mu}}+ \boldsymbol{r}
(\boldsymbol{x},t)' {\bf R}_{\bf{X},t}^{-1}(\boldsymbol{y}-{\bf
V}\hat{\boldsymbol{\mu}}),
\end{equation}
where ${\bf V}=(\boldsymbol{\upsilon}({{\bf X}_1,t^*_1}),\cdots,\boldsymbol{\upsilon}({\bf X}_N,t^*_N))'$,
$\hat{\boldsymbol{\mu}}=({\bf
V}'{\bf R}_{\bf{X},t}^{-1}{\bf V})^{-1} {\bf V}' {\bf R}_{\bf{X},t}^{-1}\boldsymbol{y}$,
$\boldsymbol{r}(\boldsymbol{x},t)=(r(\boldsymbol{x}-{\bf X}_1,
t-t^*_1),\cdots, r(\boldsymbol{x}-{\bf X}_N,t-t^*_N))'$, and ${\bf R}_{\bf{X},t}$
is the $N\times N$ correlation matrix with elements $r({\bf X}_i-{\bf
X}_j,t^*_i-t^*_j)$. Note that, the kriging  predictor in (\ref{KRIGPRED}) interpolates the data because $\hat{y}(\boldsymbol{x}_{i},t_{ij})=
\boldsymbol{\upsilon}(\boldsymbol{x}_i,t_{ij})'\hat{\boldsymbol{\mu}}+ (y_{ij}-\boldsymbol{\upsilon}(\boldsymbol{x}_i,t_{ij})'\hat{\boldsymbol{\mu}})=y_{ij}$.
The correlation parameters are estimated by
minimizing the negative log-likelihood (Santner et al. 2003, pp. 66)
 \begin{equation}\label{thetahat}
\hat{\boldsymbol{\xi}}=\mbox{arg} \min_{\boldsymbol{\xi}} \bigg{[}N \mbox{log}
\hat{\sigma}^2+\mbox{log} |{\bf R}_{\bf{X},t}|\bigg{]},
\end{equation}
where $ \hat{\sigma}^2=\frac{1}{N}(\boldsymbol{y}-{\bf
V}\hat{\boldsymbol{\mu}})'{\bf R}_{\bf{X},t}^{-1}(\boldsymbol{y}-{\bf
V}\hat{\boldsymbol{\mu}})$. Other approaches, such as cross-validation and restricted maximum likelihood (Santner et al. 2003, Section 3.3) can also be applied.

As mentioned before, an optimization algorithm employed in (\ref{thetahat})  may require hundreds of evaluations of ${\bf R}_{\bf{X},t}^{-1}$ and $|{\bf R}_{\bf{X},t}|$, which makes the estimation of correlation parameters computationally prohibitive. To tackle
the computational difficulty associated with functional response
and successfully extend kriging, we propose a new procedure in the next section.\\

\begin{center}
{\large\bf 3. Analysis of Functional Response}
\end{center}

{\large\bf 3.1 The first stage}

We propose to build the kriging model in two stages. In the first stage, we use the marginal profiles in $\bm x$ and $t$ to identify the functional form of the mean $\bm v(\bm x,t)$ and obtain the initial estimates of the correlation parameters. It can be described as follows. Consider the model
\begin{equation}\label{stage2}
y(\boldsymbol{x},t)=\mu_0+ {\bm k}'(t)\boldsymbol{u}_t+{\bm g}'(\boldsymbol{x}) \boldsymbol{\nu}_{\boldsymbol{x}}+Z(\boldsymbol{x},t),
\end{equation}
where ${\bm k}(t)=(k_1(t),\cdots,k_a(t))'$ and ${\bm g}(\boldsymbol{x})=(g_1(\boldsymbol{x}),\cdots,g_b(\boldsymbol{x}))'$. This is a special case of (\ref{OK2}) because the mean does not entertain the possible interactions between $\bm x$ and $t$. These interactions are now absorbed into $Z(\bm x,t)$.

First, we compute the average profile. Assume that the functional outputs are collected from $m$ different locations of the functional space. If the functional data are collected over an irregular grid, $m$ is defined to be the union of the abscissae of the individual profiles.
We assume that there are $n_j$ observations for location $j$, $j=1,\cdots,m$. For each $j$, define $\bar{e}_{\cdot j}=n_j^{-1}\sum_{i=1}^{n_j} (y_{ij}-\bar{y}_{i\cdot})$, where $\bar{y}_{i\cdot}=m_i^{-1}\sum_{l=1}^{m_i} y_{il}$. Note that we removed the effect of $x$-variables by subtracting $\bar{y}_{i.}$ from $y_{ij}$. Based on $\bar{e}_{\cdot 1},\cdots, \bar{e}_{\cdot m}$, we fit the model
\begin{equation}\label{meant}
\bar{e}(t)=\mu_{t0}+{\bm k}'(t)\boldsymbol{u}_t+Z(t).
\end{equation}
Next, we compute the average responses over the functional space, i.e., $\bar{y}_{1\cdot},\cdots,\bar{y}_{n \cdot}$, and fit the model
\begin{equation}\label{meanx}
\bar{y}(\boldsymbol{x})=\mu_{\boldsymbol{x}0}+{\bm g}'(\boldsymbol{x})\boldsymbol{\nu}_{\boldsymbol{x}}+Z(\boldsymbol{x}).
\end{equation}
The unknown functions in $\bm k(t)$ and $\bm g(\bm x)$ can be identified using a variable selection procedure, such as the Bayesian forward selection procedure implemented in blind kriging (Joseph et al. 2008). For the correlation functions, we consider the widely used power exponential correlation
functions $ r_i(x_{i1}-x_{i2})=\exp\{-\alpha_i
|x_{i1}-x_{i2}|^d\}$ for $\bm x$ and $ r_T(t_{1}-t_{2})=\exp\{-\beta
|t_{1}-t_{2}|^d\}$ for $t$, where $d=1$ and $d=2$ correspond to
the exponential and Gaussian correlation functions, respectively. The estimates of the
correlation parameters are denoted by $\boldsymbol{\alpha}^{(0)}=(\alpha_1^{(0)},\cdots,\alpha_p^{(0)})'$ and $\beta^{(0)}$, which are used as initial values in the second stage estimation.

The second stage, described in Sections 3.2 and 3.3, is implemented depending on how the functional outputs are collected.

{\large\bf 3.2 The second stage: regular grid}

Suppose the functional output is collected over a regular grid, i.e., ${\bf t}_1=\cdots={\bf t}_n={\bf t}$ and $m_1=\cdots=m_n=m$.
Thus, in this situation, the functional
outputs are observed at the same locations of the functional space for each experimental setting. It follows that
${\bf R}_{\bf{X},t}={\bf R}_{\bf{X}}\otimes {\bf R}_{\bf t}$, where ${\bf R}_{\bf{X}}$ is a
$n$-by-$n$ correlation matrix corresponding to $\boldsymbol{x}$ with elements $r(\boldsymbol{x}_i-\boldsymbol{x}_j)$
and ${\bf R}_{\bf t}$ is a $m$-by-$m$ correlation matrix according to the ${\bf t}$ components (Williams et al. 2006, Rougier 2008, Liu et al. 2008, Bayarri et al. 2009). We have ${\bf R}_{\bf{X},t}^{-1}={\bf R}_{\bf{X}}^{-1}\otimes {\bf R}_{\bf t}^{-1}$.
Consequently, the computational complexity involved in finding the inverse of ${\bf R}_{\bf{X},t}$ is dramatically reduced from $O(n^3m^3)$ to $O(n^3+m^3)$ (An and Owen 2001).
The resulting kriging predictor can be written as
\begin{equation}\label{KrigPredNew}
\hat{y}(\boldsymbol{x},t)=
\boldsymbol{\upsilon}(\boldsymbol{x},t)'\hat{\boldsymbol{\mu}}+ \boldsymbol{r}
(\boldsymbol{x},t)'
\bigg{(}{\bf R}_{\bf{X}}^{-1}\otimes {\bf R}_{\bf t}^{-1}\bigg{)}(\boldsymbol{
y}-{\bf V}\hat{\boldsymbol{\mu}}),
\end{equation}
where $\boldsymbol{\upsilon}(\boldsymbol{x},t)=(1,{\bm k}'(t),{\bm g}'(\boldsymbol{x}))'$ and ${\bf V}=(\boldsymbol{\upsilon}({{\bf X}_1,t^*_1}),\cdots,\boldsymbol{\upsilon}({\bf X}_N,t^*_N))'$.
The maximum likelihood estimate (MLE) of the parameters are given by
\begin{equation}\label{muhatnew}
\hat{\boldsymbol{\mu}}=\bigg{[}{\bf
V}'\bigg{(} {\hat{\bf R}}_{\bf{X}}^{-1}\otimes {\hat{\bf R}}_{\bf t}^{-1}\bigg{)}{\bf
V}\bigg{]}^{-1}\bigg{[}{\bf V}'
\bigg{(} {\hat{\bf R}}_{\bf{X}}^{-1}\otimes {\hat{\bf R}}_{\bf t}^{-1}\bigg{)}\boldsymbol{
y}\bigg{]},
\end{equation}
\begin{equation}\label{sigmahatnew}
\hat{\sigma}^2=\frac{1}{N}({\bf y}-{\bf
V}\hat{\boldsymbol{\mu}})'\bigg{(} {\hat{\bf R}}_{\bf{X}}^{-1}\otimes {\hat{\bf R}}_{\bf t}^{-1}\bigg{)}(\boldsymbol{
y}-{\bf V}\hat{\boldsymbol{\mu}}),
\end{equation}
\begin{equation}\label{thetahatnew}
\hat{\boldsymbol{\xi}}=\mbox{arg}\min_{\boldsymbol{\xi}} \bigg{[} N \mbox{log}
\hat{\sigma}^2+m\mbox{log} | {\bf R}_{\bf{X}}|+n \mbox{log}
| {\bf R}_{\bf t}| \bigg{]}.
\end{equation}

The computational complexity can be further reduced using an idea in Finley et al. (2009), provided the grids are equally spaced. This is especially useful when the functional responses are collected with intensive sampling rate. Without loss of generality, assume that the equal spacing functional outputs are observed at locations ${\bf t}=(1, 2,\cdots, m)$, where $m$ can be extremely large. In this case, the evaluation of the kriging predictor can be simplified by using an exponential correlation function ($d=1$). The use of exponential correlation function results in a non-differentiable predictor, however, lower degrees of smoothness can still be maintained for large $m$. The
correlation matrix corresponding to ${\bf t}$ can be written as
\begin{equation}\label{Requal}
{\bf R}_{\bf t}=\left[\begin{array}{ccccc}
1 &\rho &\rho^2&\cdots&\rho^m\\
\rho & 1& \rho&\cdots & \rho^{m-1}\\
\vdots& \vdots& \vdots& & \vdots\\
\rho^m & \rho^{m-1} & \rho^{m-2}&\cdots & 1\\
\end{array}
\right],
\end{equation}
where $\rho=\exp(-\beta)$. It follows that $|{\bf R}_{\bf t}|=(1-\rho^2)^m$ and
$$
{\bf R}_{\bf t}^{-1}=\frac{1}{1-\rho^2}\left[\begin{array}{ccccc}
1 &-\rho &0&\cdots&0\\
-\rho & 1+\rho^2 & -\rho &\cdots & 0\\
\vdots & \vdots & \vdots & & \vdots\\
0 & 0 & -\rho &1+\rho^2 & -\rho\\
0 & 0 &0 &-\rho & 1
\end{array}
\right].
$$
Because ${\bf R}_{\bf t}^{-1}$ is highly sparse and $|{\bf R}_{\bf t}|$
can be written in a simple closed form, the evaluation of kriging predictor
(\ref{KrigPredNew}) becomes much easier with a computational complexity of $O(n^3+mn^2)$.

{\large\bf 3.3 The second stage: irregular grid}

Consider a more general situation where the functional responses are collected over an irregular grid. That is, for the $i$th experimental setting, functional response $\boldsymbol{y}_i$ is observed at ${\bf t}_i=(t_{i1},\cdots,t_{im_i})'$, where
${\bf t}_i$'s and the lengths $m_i$'s  are not necessarily the same for every $i$, $i=1,\cdots,n$. In this situation, the Kronecker product technique, which is the key to simplify the computation, cannot be directly utilized. We propose an intuitive idea to overcome the computational problem, which is to fill-in the ``missing observations'' so that the data appear to be on a regular grid and apply the procedure developed in the previous subsection.
Even though the issue of missing data has been extensively studied in the literature, the idea of creating such missing observations to overcome the computational difficulty in kriging  is new. Moreover, a direct application of the Expectation Maximization (EM) algorithm (Dempster et al. 1977)  does not help in reducing the computational complexity. Here we introduce a carefully constructed iterative algorithm within the framework of EM algorithm to efficiently estimate the missing observations.

Define the missing data by ${\bm z}=(\bm z_1,\cdots,\bm z_n)'$, where $\bm z_i$ is a vector of missing observations in the $i$th run. Combining ${\bm z}$ and the observed data $\boldsymbol{y}=(\boldsymbol{y}'_1,\cdots, \boldsymbol{y}'_n)'$, we obtain the complete dataset $\bm c=(\bm c'_1,\cdots, \bm c'_n)'$ with $\bm c_i=(\boldsymbol{y}'_i\cup \bm z'_i)'$, which can be viewed as the data collected on a regular grid  ${\bf t}=(t_1,\cdots t_m)'$ with missing data $\bm z_i$ on ${\bf t} \backslash {\bf t}_i$. The E-step in the EM algorithm is to obtain
\begin{equation}\label{Qeq}
Q(\boldsymbol{\theta}|\hat{\boldsymbol{\theta}}^{(k)})=E(l_{\bm c}(\boldsymbol{\theta})|\boldsymbol{y}, \hat{\boldsymbol{\theta}}^{(k)}),
\end{equation}
where
\begin{equation}\label{likelihood}
l_{\bm c}(\boldsymbol{\theta})=\frac{N}{2} log(2\pi)+\frac{N}{2}log(\sigma^2)+\frac{1}{2} log| {\bf R}_{\boldsymbol{x},t}|+\frac{1}{2\sigma^2} (\boldsymbol{c}-\boldsymbol{V}\boldsymbol{\mu})' {\bf R}_{\bf{X},t}^{-1}(\boldsymbol{c}-\boldsymbol{V}\boldsymbol{\mu})
\end{equation}
is the negative log-likelihood of the complete data and $\hat{\boldsymbol{\theta}}^{(k)}=(\hat{\boldsymbol{\mu}}^{(k)}, \hat{\sigma}^{2(k)}, \hat{\boldsymbol{\xi}}^{(k)})$ are the parameters estimated at the $k$th EM iteration.
Clearly, the E-step in (\ref{Qeq}) requires the computation of
\begin{equation}\label{oldEstep}
\begin{array}{rl}
E((\boldsymbol{c}-\boldsymbol{V}\boldsymbol{\mu})' {\bf R}_{\bf{X},t}^{-1}(\boldsymbol{c}-\boldsymbol{V}\boldsymbol{\mu})|\boldsymbol{y}, \hat{\boldsymbol{\theta}}^{(k)})=&\big{(}E(\boldsymbol{c}|\boldsymbol{y}, \hat{\boldsymbol{\theta}}^{(k)})-\boldsymbol{V}\boldsymbol{\mu}\big{)}' {\bf R}_{\bf{X},t}^{-1} \big{(}E(\boldsymbol{c}|\boldsymbol{y}, \hat{\boldsymbol{\theta}}^{(k)})-\boldsymbol{V}\boldsymbol{\mu}\big{)}\\
&+\mbox{tr}\big{(}{\bf R}_{\bf{X},t}^{-1} \mbox{cov}(\boldsymbol{c}|\boldsymbol{y}, \hat{\boldsymbol{\theta}}^{(k)})\big{)},
\end{array}
\end{equation}
which cannot be done efficiently because the expressions of
$E(\bm c|\boldsymbol{y}, \hat{\boldsymbol{\theta}}^{(k)})$ and $\mbox{cov}(\boldsymbol{c}|\boldsymbol{y}, \hat{\boldsymbol{\theta}}^{(k)})$ contain the inverse of the correlation matrix constructed using the observations on an irregular grid. This is the same computational hindrance we faced before. Therefore, such a direct application of EM cannot overcome the computational issue.

To simplify the computation, we modify the standard EM approach using an idea similar to Monte Carlo EM algorithm (Chan and Ledolter 1995).
The main idea is to estimate the missing data run-by-run by conditioning on the complete data in the other runs. Let $\bm z_i^j$ be the missing data estimated in the $j$th iteration and $\bm c^j=(\bm c_1^j,\cdots, \bm c_n^j)$, where $\bm c_i^j=(\boldsymbol{y}_i \cup\bm z_i^j)$ are interspersed onto the regular grid ${\bf t}$. We can sample the missing data by using a  Gibbs sampler according to the conditional distributions
\begin{equation}\label{gibbs2}
\begin{array}{rcl}
\bm c_1^{j} &\sim & f(\bm c_1|\bm y, \bm z_{(-1)}^{j-1}, \hat{\boldsymbol{\theta}}^{(j-1)}), \\
\bm c_2^{j} &\sim & f(\bm c_2|\bm y, \bm z_1^{j}, \bm z_{(-1,-2)}^{j-1},\hat{\boldsymbol{\theta}}^{(j-1)}), \\
\vdots & & \vdots\\
\bm c_n^{j}&\sim& f(\bm c_n|\bm y, \bm z_1^{j},\cdots \bm z_{n-1}^{j},
\hat{\boldsymbol{\theta}}^{(j-1)}),
\end{array}
\end{equation}
with $\bm z^{j}_{(-i)}=(\bm z^{j}_1,\cdots,\bm z^{j}_{i-1}, \bm z^{j}_{i+1},\cdots, \bm z^{j}_n)$. Because each step in (\ref{gibbs2}) involves data from a regular grid, the computations are cheap. This is shown below.

The prior distribution (Currin et al. 1991) for ${\bm c}_i$ is
\begin{equation}\label{prior1}
f(\bm c_i|\hat{\boldsymbol{\theta}}^{(k)}) \sim N\big{(}\boldsymbol{\zeta}_{\bm c_i}, {\bf \Sigma}_{\bm c_i}\big{)},
\end{equation}
where $\boldsymbol{\zeta}_{\bm c_i}=\big{(}\boldsymbol{\upsilon}(\boldsymbol{x}_i, t_1),\cdots, \boldsymbol{\upsilon}(\boldsymbol{x}_i, t_m)\big{)}'\boldsymbol{\mu}$ and ${\bf \Sigma}_{\bm c_i}=\sigma^2 {\bf R}_{\bf t}$.
It follows that
\begin{equation}\label{prior2}
f(\bm c_i|\boldsymbol{y}_{i},\hat{\boldsymbol{\theta}}^{(k)})\sim N\big{(}\boldsymbol{\zeta}_{i}, {\bf \Sigma}_{i}\big{)},
\end{equation}
\begin{equation}\label{prior3}
f(\bm c_i|\boldsymbol{y}_{(-i)}, \bm z_{(-i)},\hat{\boldsymbol{\theta}}^{(k)})\sim N \big{(}\boldsymbol{\zeta}_{(-i)},{\bf \Sigma}_{(-i)}\big{)},
\end{equation}
where
$$
\begin{array}{rl}
\boldsymbol{\zeta}_{i}=&\boldsymbol{\zeta}_{\bm c_i}+(\boldsymbol{r}_{{\bf t}_i}(t_1),\cdots, \boldsymbol{r}_{{\bf t}_i}(t_m))' {\bf R}_{{\bf t}_i}^{-1}\big{[}\boldsymbol{y}_{i}-\big{(}\boldsymbol{\upsilon}(\boldsymbol{x}_i, t_{i1}),\cdots, \boldsymbol{\upsilon}(\boldsymbol{x}_i, t_{im_i})\big{)}'\boldsymbol{\mu}\big{]},\\
{\bf \Sigma}_{i}=&{\bf \Sigma}_{\bm c_i}-\sigma^2(\boldsymbol{r}_{{\bf t}_i}(t_1),\cdots,\boldsymbol{r}_{{\bf t}_i}(t_m))' {\bf R}_{{\bf t}_i}^{-1}(\boldsymbol{r}_{{\bf t}_i}(t_1),\cdots,\boldsymbol{r}_{{\bf t}_i}(t_m)),\\
\boldsymbol{\zeta}_{(-i)}=&\boldsymbol{\zeta}_{\bm c_i}+\big{(}\boldsymbol{r}_{(-i)}(\boldsymbol{x}_{i},t_1),\cdots,\boldsymbol{r}_{(-i)}(\boldsymbol{x}_{i},t_m)\big{)}'\bigg{(}{\bf R}_{\bf{X}_{(-i)}}^{-1}\otimes {\bf R}^{-1}_{\bf t}\bigg{)}\big{(}\bm c_{(-i)}-\boldsymbol{\zeta}_{\bm c_{(-i)}}\big{)},\\
{\bf \Sigma}_{(-i)}=&{\bf \Sigma}_{\bm c_i}-\sigma^2 \big{(}\boldsymbol{r}_{(-i)}(\boldsymbol{x}_{i},t_1),\cdots,\boldsymbol{r}_{(-i)}(\boldsymbol{x}_{i},t_m)\big{)}'\\
&\bigg{(}{\bf R}^{-1}_{\bf{X}_{(-i)}}\otimes {\bf R}^{-1}_{\bf t}\bigg{)}\big{(}\boldsymbol{r}_{(-i)}(\boldsymbol{x}_{i},t_1),\cdots,\boldsymbol{r}_{(-i)}(\boldsymbol{x}_{i},t_m)\big{)},
\end{array}
$$
$\boldsymbol{r}_{{\bf t}_i}(t_j)=(r_T(t_j-t_{i1}),\cdots,r_T(t_j-t_{im_i}))'$, $\boldsymbol{\zeta}_{\bm c_{(-i)}}=E(\bm c_{(-i)})$,
${\bf R}_{\bf{X}_{(-i)}}$ is the $(n-1)\times (n-1)$ correlation matrix corresponding to $\boldsymbol{x}$ except
the $i$th setting,
$\boldsymbol{y}_{(-i)}$ is the observed data except the $i$th profile, $\bm z_{(-i)}$ is the missing data except those from the $i$th run, and
$\bm c_{(-i)}=(\boldsymbol{y}_{(-i)} \cup \bm z_{(-i)})$.
Then, we obtain the following result.

\noindent\textbf{Proposition 1:}  Based on (\ref{prior1}), (\ref{prior2}), and (\ref{prior3}), it follows that
\begin{equation}\label{post}
f(\bm c_i|\boldsymbol{y}, \bm z_{(-i)},\hat{\boldsymbol{\theta}}^{(k)})\sim N(\boldsymbol{\eta}_i, \boldsymbol{\Gamma}_i),
\end{equation}
where
\begin{equation}\label{pmean}
\boldsymbol{\eta}_i=({\bf \Sigma}_{(-i)}^{-1}+{\bf \Sigma}_i^{-1}-{\bf \Sigma}^{-1}_{\bm c_i})^{-1}({\bf \Sigma}_i^{-1}\boldsymbol{\zeta}_i+{\bf \Sigma}_{(-i)}^{-1}\boldsymbol{\zeta}_{(-i)}-{\bf \Sigma}^{-1}_{\bm c_i}\boldsymbol{\zeta}_{\bm c_i}),
\end{equation}
\begin{equation}\label{pvariance}
\boldsymbol{\Gamma}_i=({\bf \Sigma}_{(-i)}^{-1}+{\bf \Sigma}_i^{-1}-{\bf \Sigma}^{-1}_{\bm c_i} )^{-1}.
\end{equation}

The proofs are given in the Appendix. Based on (\ref{post}), the conditional mean of $\bm c_i$ is a weighted average of three terms: the mean $\boldsymbol{\zeta}_{\bm c_i}$ from prior distribution, the conditional mean based on the $i$th observed profile $\boldsymbol{\zeta}_i=E(\bm c_i|\boldsymbol{y}_{i},\hat{\boldsymbol{\theta}}^{(k)})$, and the conditional mean based on the complete data except the $i$th profile $\boldsymbol{\zeta}_{(-i)}=E(\bm c_i|\bm c_{(-i)},\hat{\boldsymbol{\theta}}^{(k)})$. The weights are determined by their corresponding variances, ${\bf \Sigma}_{\bm c_i}$, ${\bf \Sigma}_i$, and ${\bf \Sigma}_{(-i)}$, which are also components in the conditional variance. Evaluating the conditional distribution in (\ref{post}) is computationally efficient because $\boldsymbol{\zeta}_{\bm c_i}$ and $\boldsymbol{\zeta}_i$ involve observations with size $m$ and $m_i$ which are easy to calculate. This is also true for the variances ${\bf \Sigma}_{\bm c_i}$ and ${\bf \Sigma}_i$. Furthermore, because $\bm c_{(-i)}$ is  on a regular grid, $\boldsymbol{\zeta}_{(-i)}$ and ${\bf \Sigma}_{(-i)}$ can be easily evaluated.
By simple modifications, results in Proposition 1 can be extended to the conditional distributions in (\ref{gibbs2}). For example, the posterior distribution $f(\bm c_i|\boldsymbol{y}, \bm z_1^{j},\bm z^{j-1}_{(-1,-2)},\hat{\boldsymbol{\theta}}^{(k)})$ can be obtained by replacing $\bm z_{(-i)}$ with the updated missing data $\big{(}\bm z_1^{j},\bm z^{j-1}_{(-1,-2)}\big{)}$ in (\ref{post}).

After generating $q$ samples using the Gibbs sampler, we can now approximate  $Q(\boldsymbol{\theta}|\hat{\boldsymbol{\theta}}^{(k)})$ in (\ref{Qeq}) by
\begin{equation}\label{Q1}
Q^{(q)}(\boldsymbol{\theta}|\hat{\boldsymbol{\theta}}^{(k)})=\frac{1}{q}\sum_{j=1}^q l_{\bm c^j}(\boldsymbol{\theta}).
\end{equation}
However, if the Gibbs sampler needs too many samples to attain convergence, the computational advantage of the foregoing procedure may diminish. Interestingly, we can further simplify the procedure and speed-up the convergence using the following idea. Instead of sampling from the conditional distributions, we can simply iterate  on the conditional expectations as follows:
\begin{equation}\label{gibbs3}
\begin{array}{rcl}
\bm z_1^{j} &= & E(\bm z_1|\bm y, \bm z_{(-1)}^{j-1}, \hat{\boldsymbol{\theta}}^{(j-1)}), \\
\bm z_2^{j} &= & E(\bm z_2|\bm y, \bm z_1^{j}, \bm z_{(-1,-2)}^{j-1},\hat{\boldsymbol{\theta}}^{(j-1)}), \\
\vdots & & \vdots\\
\bm z_n^{j} &= & E(\bm z_n|\bm y, \bm z_1^{j},\cdots \bm z_{n-1}^{j},
\hat{\boldsymbol{\theta}}^{(j-1)}),
\end{array}
\end{equation}
where $E(\bm z_i|\boldsymbol{y},\bm z_{(-i)},\hat{\boldsymbol{\theta}}^{(k)})$ represents the expectation based on the conditional distribution $f(\bm c_i|\boldsymbol{y},\bm z_{(-i)}, \hat{\boldsymbol{\theta}}^{(k)})$. These conditional expectations can be easily computed using (\ref{pmean}).
It is worth noting that
$$
{\bf R}_{\bf{X}}= \left[\begin{array}{cc} {\bf R}_{{\bf X}_{(-i)}} &{\boldsymbol \varrho}_i\\{\boldsymbol \varrho}_i' &1
\end{array}\right]
$$
and
\begin{equation}\label{update}
{\bf R}^{-1}_{\bf{X}}
=\left[\begin{array}{cc}
{\bf R}^{-1}_{{\bf X}_{(-i)}}+{\bf R}^{-1}_{{\bf X}_{(-i)}}{\boldsymbol{\varrho}}_i{\boldsymbol \varrho}_i'{\bf R}^{-1}_{{\bf X}_{(-i)}} s_R^{-1} &
-{\bf R}^{-1}_{{\bf X}_{(-i)}}{\boldsymbol \varrho}_is_R^{-1}\\
-{\boldsymbol \varrho}_i'{\bf R}^{-1}_{{\bf X}_{(-i)}}s_R^{-1} & s_R^{-1}
\end{array} \right]=\left[\begin{array}{cc}
{\bf A}_{(-i)}& {\bf a}_i\\
{\bf a}'_i & b_i
\end{array} \right],
\end{equation}
where $s_R=1-{\bf u}_i' {\bf R}^{-1}_{{\bf X}_{(-i)}} {\boldsymbol \varrho}_i$.
Therefore, the evaluation of ${\bf R}^{-1}_{{\bf X}_{(-i)}}$ in (\ref{pmean}) can be efficiently updated by
$$
{\bf R}^{-1}_{{\bf X}_{(-i)}}={\bf A}_{(-i)}-\frac{{\bf a}_i{\bf a}_i'}{b_i}.
$$
Because of this simplification, the complexity in estimating the missing observations is successfully reduced from $O\big{(}(\sum_i^n m_i)^3\big{)}$ to $O(n^3+\sum_i^n m_i^3+m^3+mn^2)$ per iteration and it can be further reduced to $O(n^3+mn^2)$ by using the exponential correlation function (\ref{Requal})  for the functional argument.
The asymptotic convergence of $\bm z_i^j$'s follows from the proposition below.

\noindent\textbf{Proposition 2:} Assume that \begin{equation}\label{prop2assump}
\mbox{max}_i \sum_{k=1,k\neq i}^n d_k \bigg{|}\bigg{|}\bigg{[}{\bm I}_{m\times m}+{\bf \Sigma}_{(-i)}\big{(}{\bf \Sigma}_i^{-1}-{\bf \Sigma}^{-1}_{\bm c_i}\big{)}\bigg{]}_{ik}^{-1}\bigg{|} \bigg{|}_2<1,
\end{equation}
where $(d_1,\cdots, d_{i-1},d_{i+1},\cdots, d_n)=\boldsymbol{r}'_{(-i)}(\boldsymbol{x}_i) {\bf R}^{-1}_{\bf{X}_{(-i)}}$, $\parallel {\bf D}\parallel_2=max_{\parallel{\bf s}\parallel_2=1} \parallel {\bf D}{\bf s}\parallel_2$, and $\big{[}\cdot\big{]}_{ik}$ indicates the block matrix with rows corresponding to the missing data at the $i$th run and columns corresponding to the missing data at the $k$th run.
Then for all $i$, $\parallel\bm z_i^q-E(\bm z_i|\bm y, \hat{\boldsymbol{\theta}}^{(k)}) \parallel_2$ converges to $\bm 0$ as $m \rightarrow \infty$ and $q \rightarrow \infty$ .

The assumption in (\ref{prop2assump}) can be interpreted intuitively. Recall that in Proposition 1 we have ${\bf \Sigma}_{i}={\bf \Sigma}_{\bm c_i}-\sigma^2(\boldsymbol{r}_{{\bf t}_i}(t_1),\cdots,\boldsymbol{r}_{{\bf t}_i}(t_m))'$ ${\bf R}_{{\bf t}_i}^{-1}(\boldsymbol{r}_{{\bf t}_i}(t_1),\cdots,\boldsymbol{r}_{{\bf t}_i}(t_m))$. When $m_i=0$, the equality ${\bf \Sigma}_{i}={\bf \Sigma}_{\bm c_i}$ holds; when $m_i\neq 0$, $|{\bf \Sigma}_{\bm c_i}-{\bf \Sigma}_{i}|>0$ and it increases with $m_i$ for fixed $\hat{\boldsymbol{\theta}}$. Similarly, it follows that $|{\bf \Sigma}_i^{-1}-{\bf \Sigma}^{-1}_{\bm c_i}|=|{\bf \Sigma}^{-1}_{\bm c_i}| |\sigma^2(\boldsymbol{r}_{{\bf t}_i}(t_1),\cdots,\boldsymbol{r}_{{\bf t}_i}(t_m))'$ ${\bf R}_{{\bf t}_i}^{-1}(\boldsymbol{r}_{{\bf t}_i}(t_1),\cdots,\boldsymbol{r}_{{\bf t}_i}(t_m))| |{\bf \Sigma}_i^{-1}|$ increases with $m_i$ and consequently the left side of (\ref{prop2assump}) decreases. Hence, this assumption suggests that the number of observations from the $i$th profile ($m_i$) should be reasonably large in order to accurately estimate the missing data.

The simplification made by (\ref{gibbs3}) results in a new approximation to the E-step:
\begin{equation}\label{Q2}
Q^{(q)}(\boldsymbol{\theta}|\hat{\boldsymbol{\theta}}^{(k)})\approx l_{\bm c^q}(\boldsymbol{\theta}),
\end{equation}
where $l_{\bm c^q}$ is the negative log-likelihood evaluated in (\ref{likelihood}) with $\bm c$ replaced by $\bm c^q=(\boldsymbol{y}, \bm z^q_1,\cdots,\bm z^q_n)$.  This is equivalent to estimating the missing observations and then pretending they were known.


Now we can implement the M-step in the EM algorithm, which is to update the MLEs by
$$
\hat{\boldsymbol{\theta}}^{(k+1)}=\mbox{argmin}_{\boldsymbol{\theta}} \,\,Q^{(q)}(\boldsymbol{\theta}|\hat{\boldsymbol{\theta}}^{(k)}),$$
where $Q^{(q)}(\boldsymbol{\theta}|\hat{\boldsymbol{\theta}}^{(k)})$ is obtained from (\ref{Q2}).  Since the estimation herein is based on the complete data $\bm c^q$ collected on regular grids, estimators according to (\ref{muhatnew}), (\ref{sigmahatnew}), and (\ref{thetahatnew})
can be computed efficiently.

There are some parameters that need to be pre-specified in the EM procedure.
The initial estimates of $\hat{\boldsymbol{\theta}}^{(0)}=(\hat{\boldsymbol{\mu}}^{(0)}, \hat{\sigma}^{2(0)}, \hat{\boldsymbol{\alpha}}^{(0)},\hat{\beta}^{(0)})$ can be obtained from the first stage of our model building procedure.
The initial estimates of the missing data can be obtained by adding the predictions from the two kriging models (\ref{meant}) and (\ref{meanx}).
Thus, the missing data of the $i$th profile, $\bm z_i^{(0)}$, at point $t$, is given by
$$
\hat{y}(\boldsymbol{x}_i,t)=\hat{\mu}_{t0}+\hat{\mu}_{x0}+{\bm k}(t)'\hat{\boldsymbol{u}}_t+{\bm g}(\boldsymbol{x}_i)'\hat{\boldsymbol{\nu}}_{\boldsymbol{x}}+\boldsymbol{r}_T(t)'{\bf R}^{-1}_{\bf t}({\boldsymbol{e}}_t-\boldsymbol{K}\hat{\boldsymbol{u}}_t-\hat{\mu}_{t0} \bm 1_{m_i})+
\boldsymbol{r}(\boldsymbol{x}_i)' {\bf R}^{-1}_{\bf{X}}(\bar{\boldsymbol{y}}_{\boldsymbol{x}}-\boldsymbol{G}\hat{\boldsymbol{\nu}}_{\boldsymbol{x}}-\hat{\mu}_{x0} \bm 1_n),
$$
where $\boldsymbol{K}=({\bm k}(t_1),\cdots,{\bm k}(t_m))'$, $\boldsymbol{G}=({\bm g}(\boldsymbol{x}_1),\cdots, {\bm g}(\boldsymbol{x}_n))'$, and ${\bf R}^{-1}_{\bf t}$ and $ {\bf R}^{-1}_{\bf{X}}$ are evaluated at $\hat{\beta}^{(0)}$ and $\hat{\boldsymbol{\alpha}}^{(0)}$.
The EM procedure is terminated when
$
\mbox{max}_i | \hat{\theta}_i^{(k+1)}-\hat{\theta}_i^{(k)}| < \Delta,
$
where the tolerance $\Delta$ can be chosen based on the required accuracy.

Since the proposed approximation converges to
the standard E-step when $q \rightarrow \infty$, the convergence of this procedure is guaranteed based on the results in the EM literature. More details can be found in Wu (1983), Chan and Ledolter (1995), and Fort and Moulines (2003).

\begin{center}
{\large\bf 4. RESIDUAL STRESS OPTIMIZATION}
\end{center}

In this section, we revisit the machining experiment originally performed by Hung et al. (2009) and apply the
proposed method to analyze the functional outputs. The objective
of the experiment is to optimize a turning process for hardened
bearing steel. Because of the computational challenge, Hung et al. (2009) optimized only the cutting forces, which is a single output problem, whereas the main objective of the experiment  was to optimize the hard turning process
with respect to residual stress, which is a functional output and is the focus of this paper.

Ten variables are considered in this experiment (Table
\ref{y:tab3}). A 30-run orthogonal-maximin  branching Latin hypercube design (BLHD) (Hung et al. 2009) is
constructed based on the first nine variables and  is
given in Table \ref{tab:data}. The cutting edge shape is labeled
``1" and ``2" to stand for chamfer and hone, respectively. The
second and third factor, chamfer angle and length, are divided
into 15 levels and the setting corresponding to the hone edge
($x_1=2$) is left blank because these two factors exist only when
chamfer edge is considered. This is the main difference between
BLHD and the traditional Latin hypercube design. The rest of the
factors are set up at 30 levels. Since the cutting edge shape is a categorical variable, an isotropic correlation function (Hung et al. 2009) is used in the analysis. Exponential correlation functions as defined in Section 3.1 are used for the rest of the factors. For each experimental setting, residual
stresses are collected as a function of depth using a highly
sophisticated finite element-based machining simulation software
\emph{AdvantEdge} (Third Wave Systems Inc., Minneapolis, MN).
The simulations are computationally intensive which
require 12 to 24 hours of running time for  each experimental
run.
\renewcommand{\baselinestretch}{1.5}
\begin{table}
\caption{Factors and their ranges in the hard turning
experiment}\label{y:tab3}
\begin{center}
\begin{tabular}{|l@{ }|c@{ }| }
\hline Factor & Ranges \\
\hline
$x_1$: Cutting edge shape & chamfer or hone \\
$x_2$: Chamfer Angle (degree) & 17 $\sim$ 20 \\
$x_3$: Chamfer Length ($\mu$m) & $115 \sim 140$ \\
$x_4$: Cutting edge radius ($\mu$m) & 5 $\sim$ 25 \\
$x_5$: Rake angle (degree) & $-15$ $\sim -5$ \\
$x_6$: Tool nose radius (mm) & $0.4$ $\sim 1.6$ \\
$x_7$: Cutting speed (m/min) & $120$ $\sim 240$ \\
$x_8$: Feed rate (mm/rev) & $0.05$ $\sim 0.15$ \\
$x_9$: Depth of cut (mm) & $0.1$ $\sim 0.25$ \\
$x_{10}$: Location & $1,2,3$\\\hline
\end{tabular}
\end{center}
\end{table}
\renewcommand{\baselinestretch}{1}
\begin{table}
\caption{Orthogonal-maximin BLHD for the hard turning
experiment}\label{tab:data}
\begin{center}
\begin{tabular}{|c|ccccccccc|}
\hline
Run & $x_1$& $x_2$ & $x_3$ &$x_4$ & $x_5$ & $x_6$ & $x_7$ & $x_8$& $x_9$ \\
\hline
1&1&1& 6& 15& 23 &7& 9 &18& 10\\
2&1&2& 11& 25& 3 &25& 14& 25& 19\\
3&1&3& 3 &4 &20 &18& 18& 5& 26 \\
4&1&4& 14& 9& 6& 6& 27& 7 &17 \\
5&1&5& 8& 16& 8 &21& 2& 2 &1 \\
6&1& 6& 1& 17& 10& 5& 25 &19& 25 \\
7&1& 7& 12& 29& 26& 15& 5 &14 &12 \\
8&1& 8& 5 &26 &16& 30& 22 &15& 6 \\
9&1& 9& 15 &7 &13& 26& 7 &11& 27 \\
10&1& 10& 10& 1& 29& 20& 23& 6 &5 \\
11&1& 11& 2& 20 &21& 27& 10& 20& 29\\
12&1& 12& 7& 8 &11& 14& 4 &29& 21\\
13&1& 13& 13& 22& 9& 1& 24& 27 &9 \\
14&1& 14& 4& 10& 2& 24& 28 &13& 13 \\
15&1& 15& 9& 28 &25& 13& 17& 3 &28 \\
16&2& & &19& 5& 9& 1& 8& 20 \\
17&2& & &14& 28& 17& 6 &21& 24 \\
18&2& & &6 &17 &4 &16 &12& 4 \\
19&2& & &11& 1 &12 &15 &4& 8 \\
20&2& & &27& 22& 8& 30 &24& 16 \\
21&2& & &21& 14& 23& 19& 10& 22\\
22&2& & &23& 18& 22& 12& 28 &3\\
23&2& & &3& 27& 3 &3 &26 &14\\
24&2& & &13& 15& 19& 29 &16& 30 \\
25&2& & &24& 12& 2& 11& 1& 18 \\
26&2& & &18& 24 &28& 8 &17& 2 \\
27&2& & &12& 30 &11& 26& 9& 11\\
28&2& & &2& 4 &16 &13 &30 &15 \\
29&2& & &30& 7 &10& 20 &23& 7 \\
30&2& & &5& 19 &29& 21 &22 &23 \\
\hline
\end{tabular}
\end{center}
\end{table}
\renewcommand{\baselinestretch}{1.5}

\begin{figure}[h]
\centering\resizebox{300pt}{250pt}
{\includegraphics{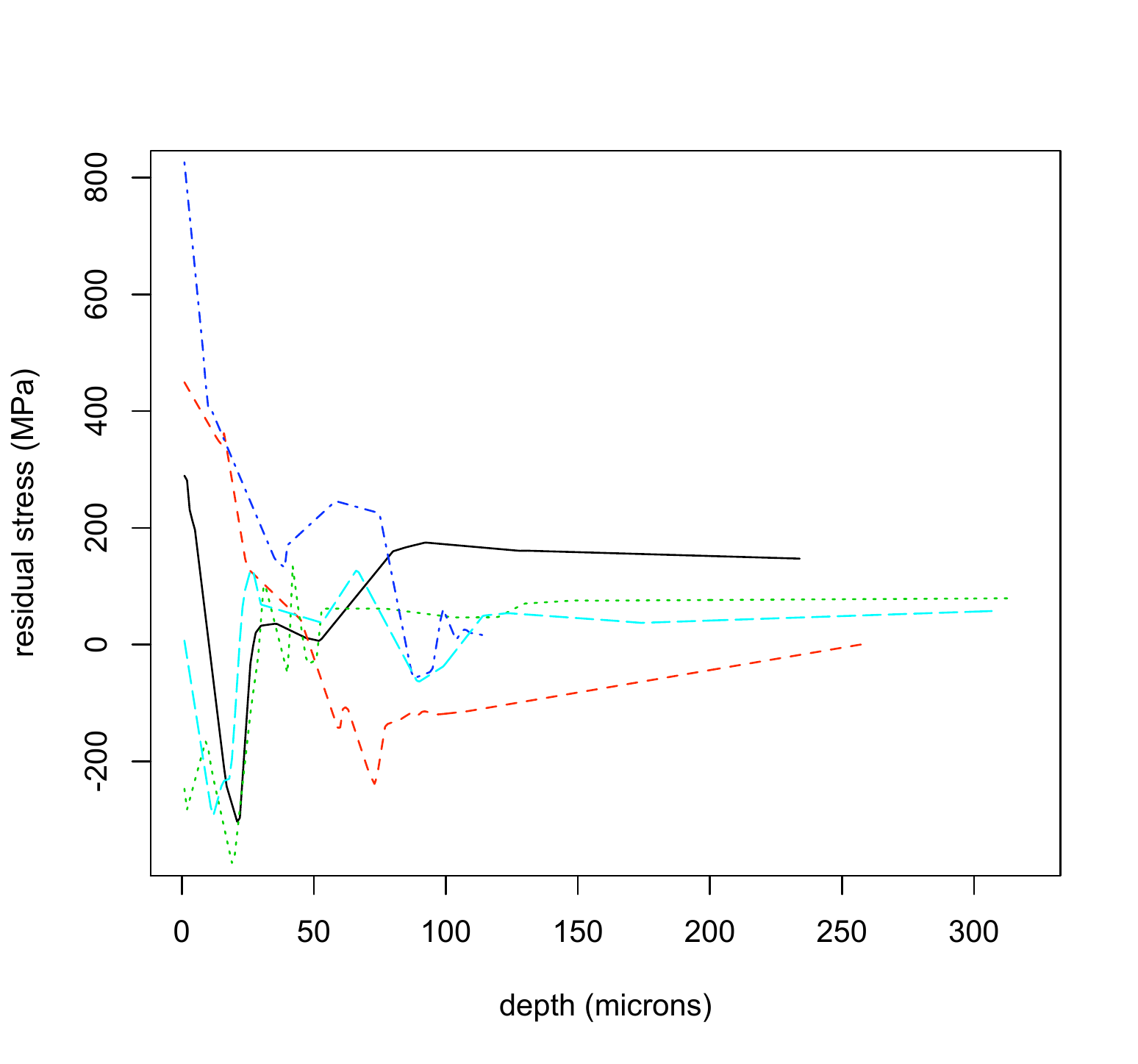}} \caption{Truncated residual stress profiles.}\label{RSnonregular}
\end{figure}

\begin{figure}[htbp]
\centering
\includegraphics[width=3.2in]{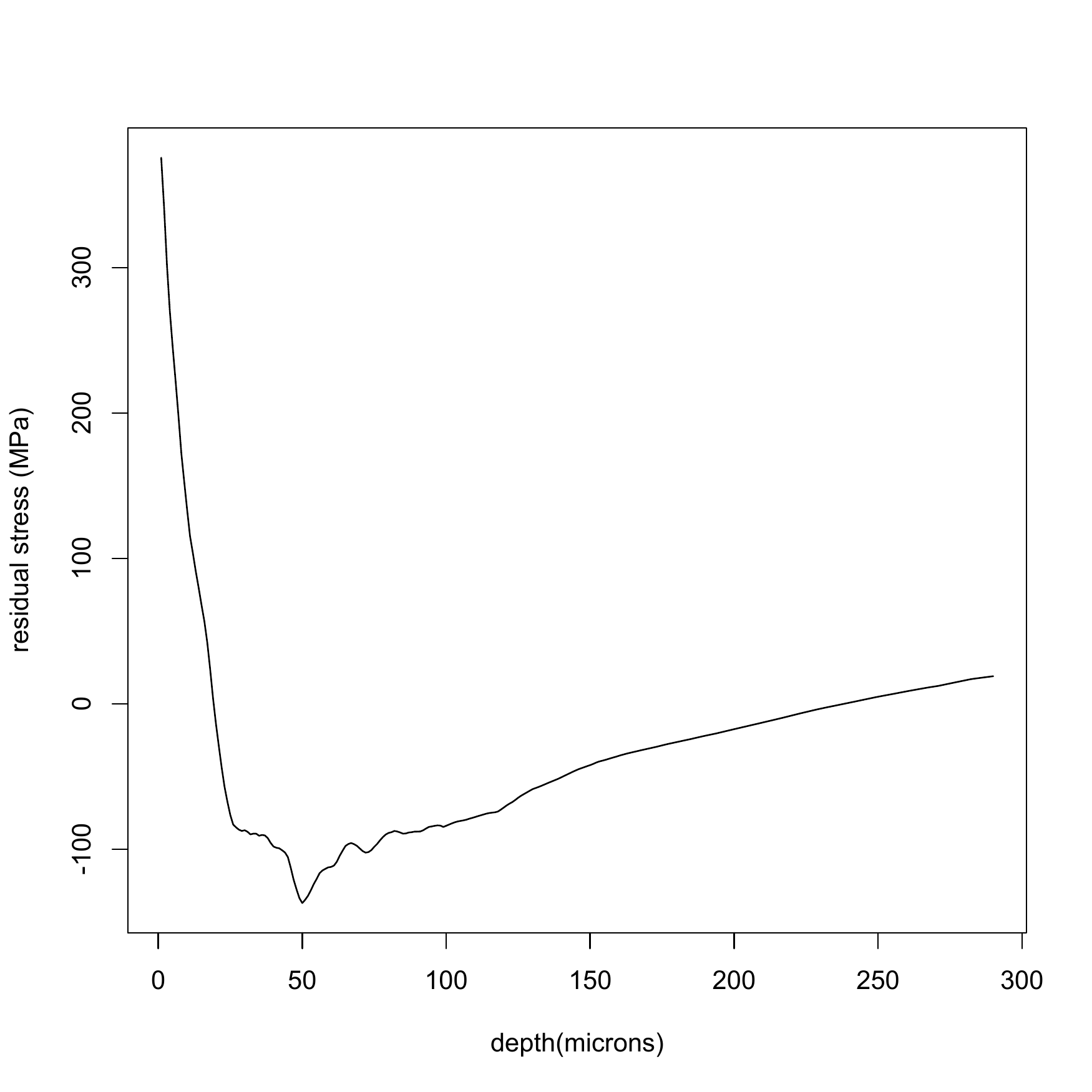}
\caption{Average residual stress profile}\label{meanprofile}
\end{figure}

The residual stress profiles are originally collected on a regular grid, but to illustrate the procedure, we artificially truncated them (randomly from 100 to 376 microns) so that the profiles are observed over an irregular grid. The assumption (\ref{prop2assump}) holds in general with such a truncation and the left hand side of (\ref{prop2assump}) decreases with respect to the percentage of truncation.  Figure \ref{RSnonregular} illustrates five such profiles.
The average of the 90 residual stress profiles  is shown in Figure \ref{meanprofile}. Theoretically, it is expected that the residual stress will tend towards 0 as the depth increases. Therefore, a nonlinear model might fit this profile better. We fitted $S=exp(-\lambda t) (\mu_0+\mu_1 t+\mu_2 t^2)$ for the averaged residual stress profile $S$ using least squares and obtained an estimate of $\lambda$ as 0.005. Based on this result, we transformed the original functional data by multiplying $exp(\lambda t)$ and used them for the model fitting.

By fitting $e(t)=\mu_{t0}+{\bm k}'(t)\boldsymbol{u}_t+Z(t)$ with $\bm k(t)=(t,t^2)'$, we obtained  $\hat{\beta}^{(0)}=1.24$. The blind kriging procedure (Joseph et al. 2008) applied on the 30 average values did not suggest any particular trend in the $x$-variables; therefore we fitted an ordinary kriging model $\bar{y}(\boldsymbol{x})=\mu_{\boldsymbol{x}0}+Z(\boldsymbol{x})$ and obtained $\hat{\boldsymbol{\alpha}}^{(0)}=(0.10, 0.76,0.45,4.50,3.72,1.63,0.17,2.05,4.50,0.59)'$. Now we are ready to move on to the second stage of the model building procedure.

\begin{figure}[htbp]
\centering
\includegraphics[width=2in,height=2.1in]{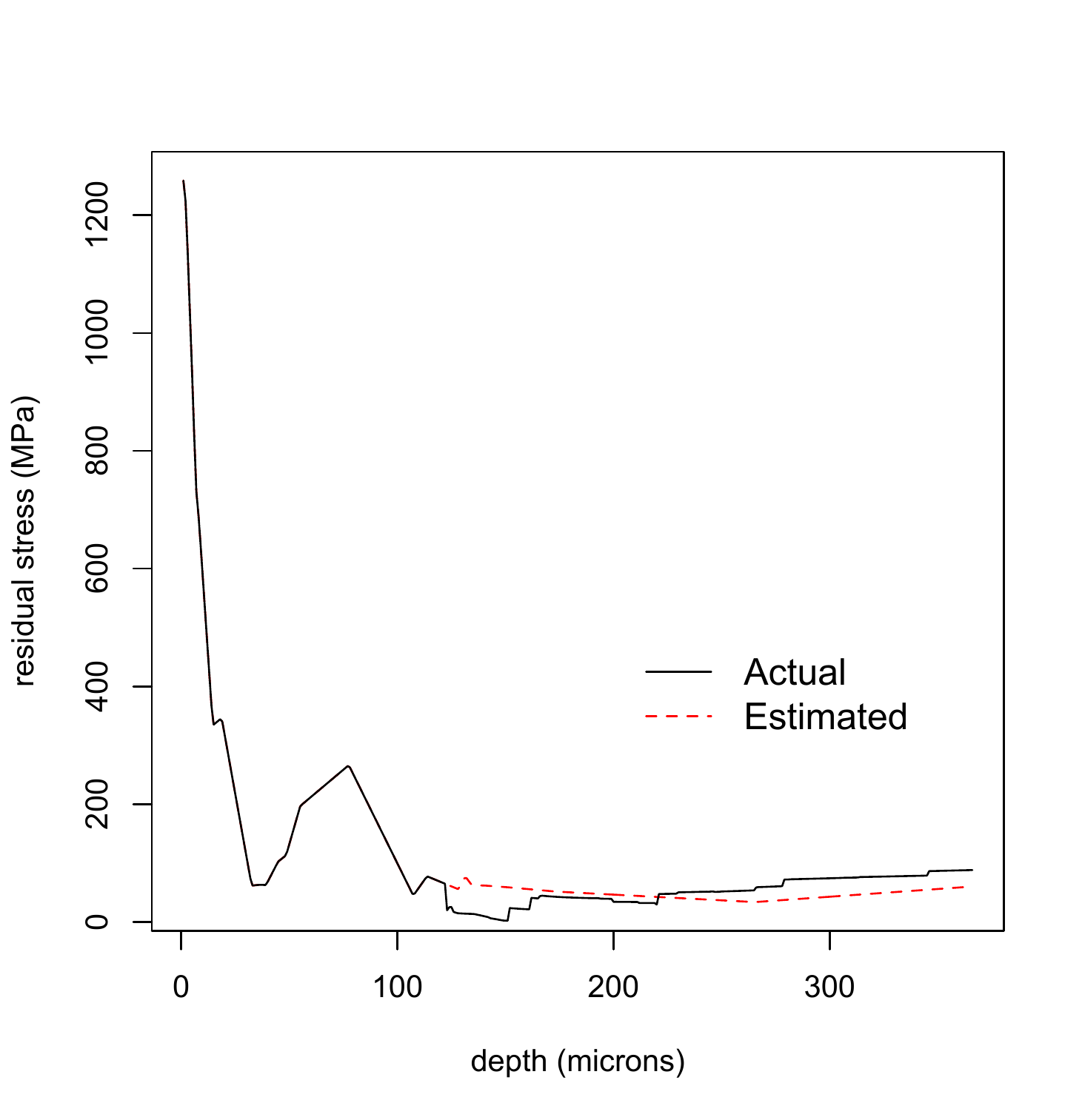}
\includegraphics[width=2in]{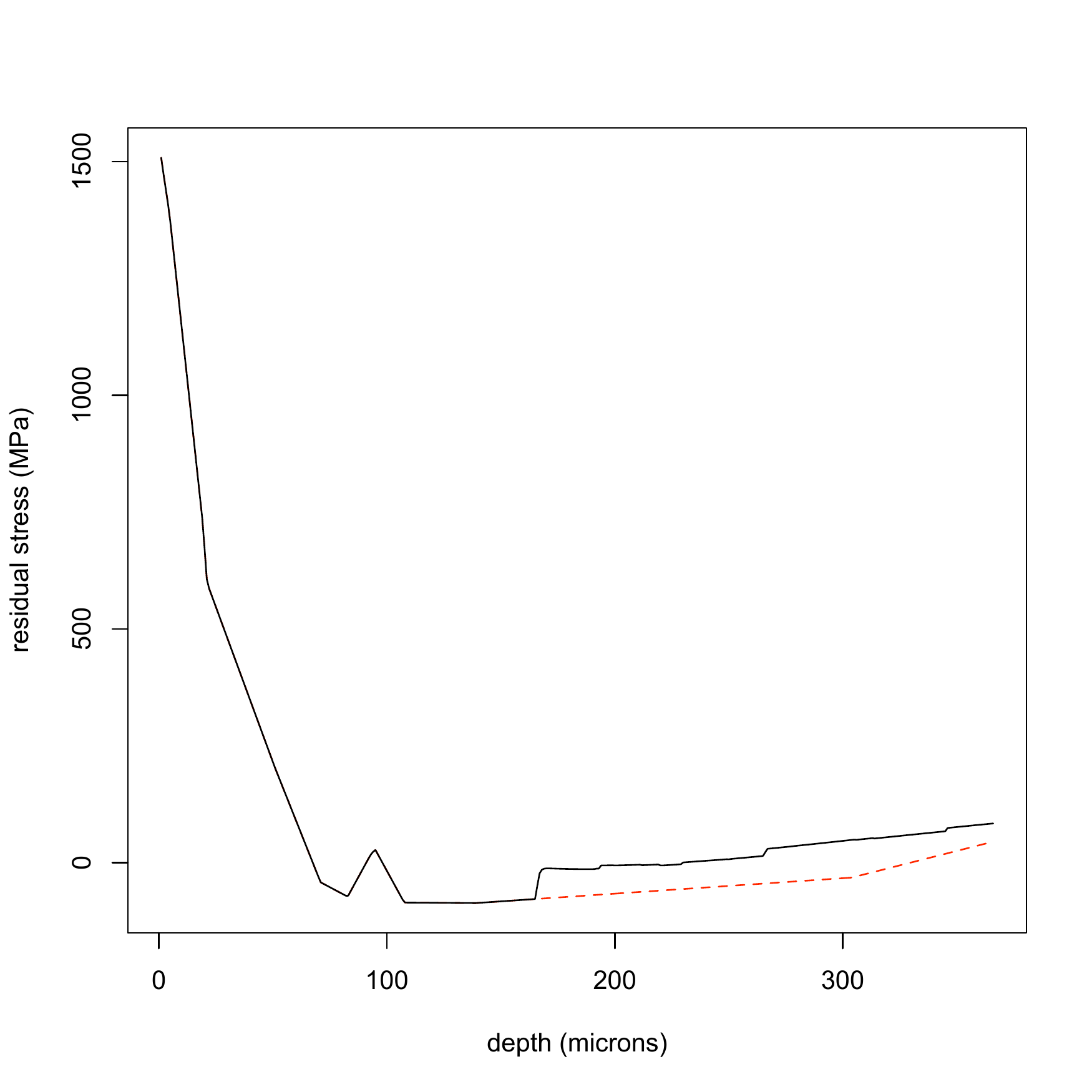}
\includegraphics[width=2in]{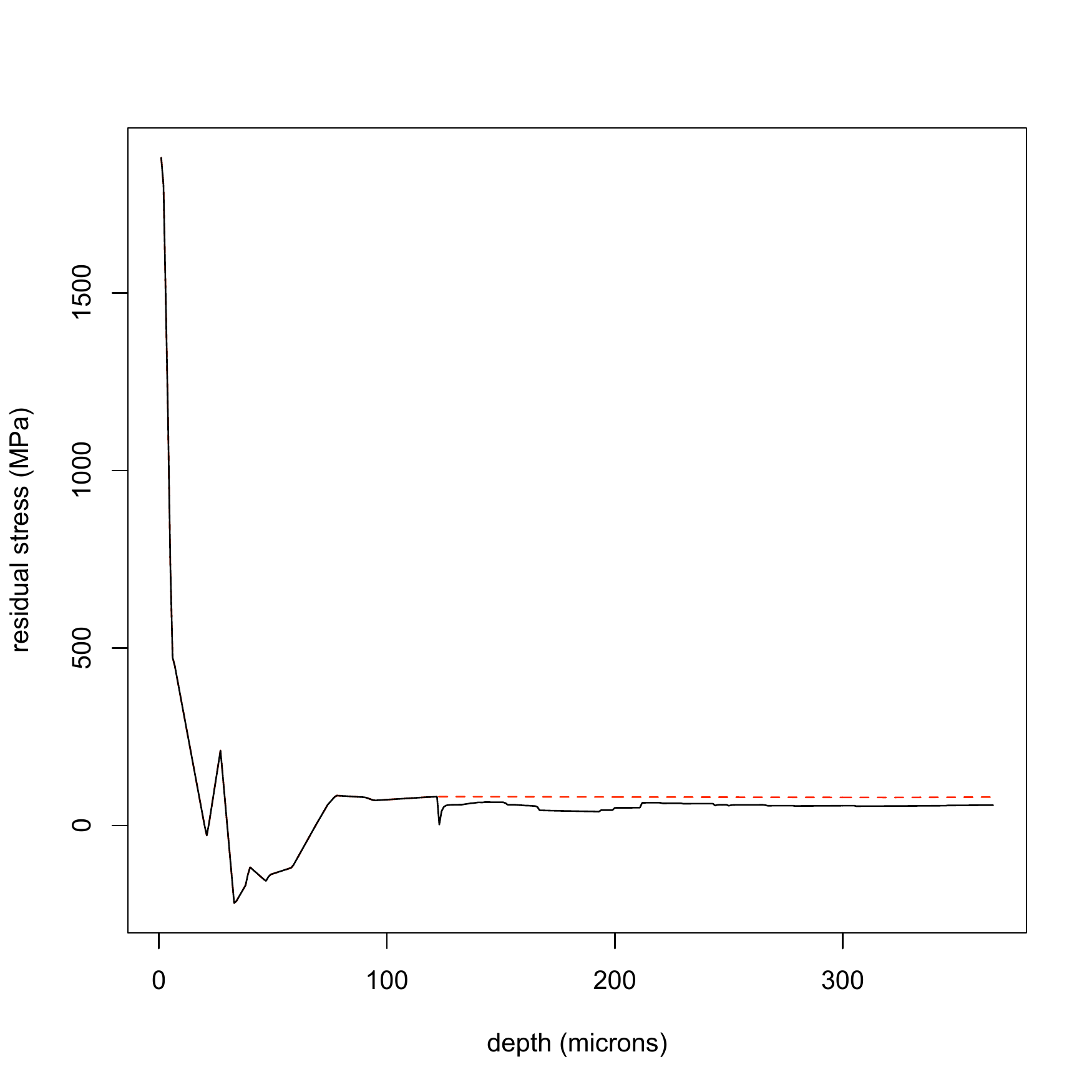}\\
\includegraphics[width=2in,height=2.1in]{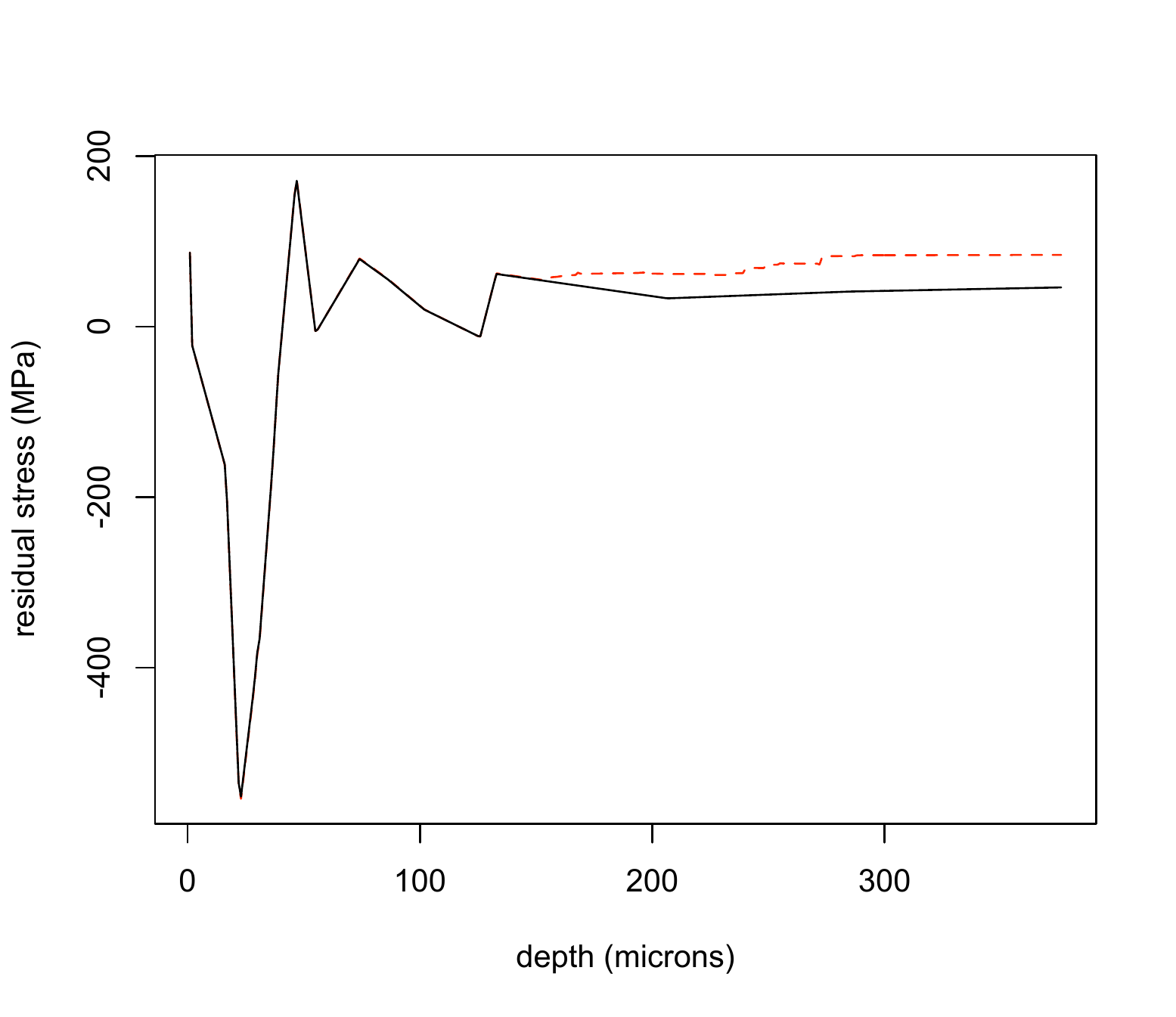}
\includegraphics[width=2in]{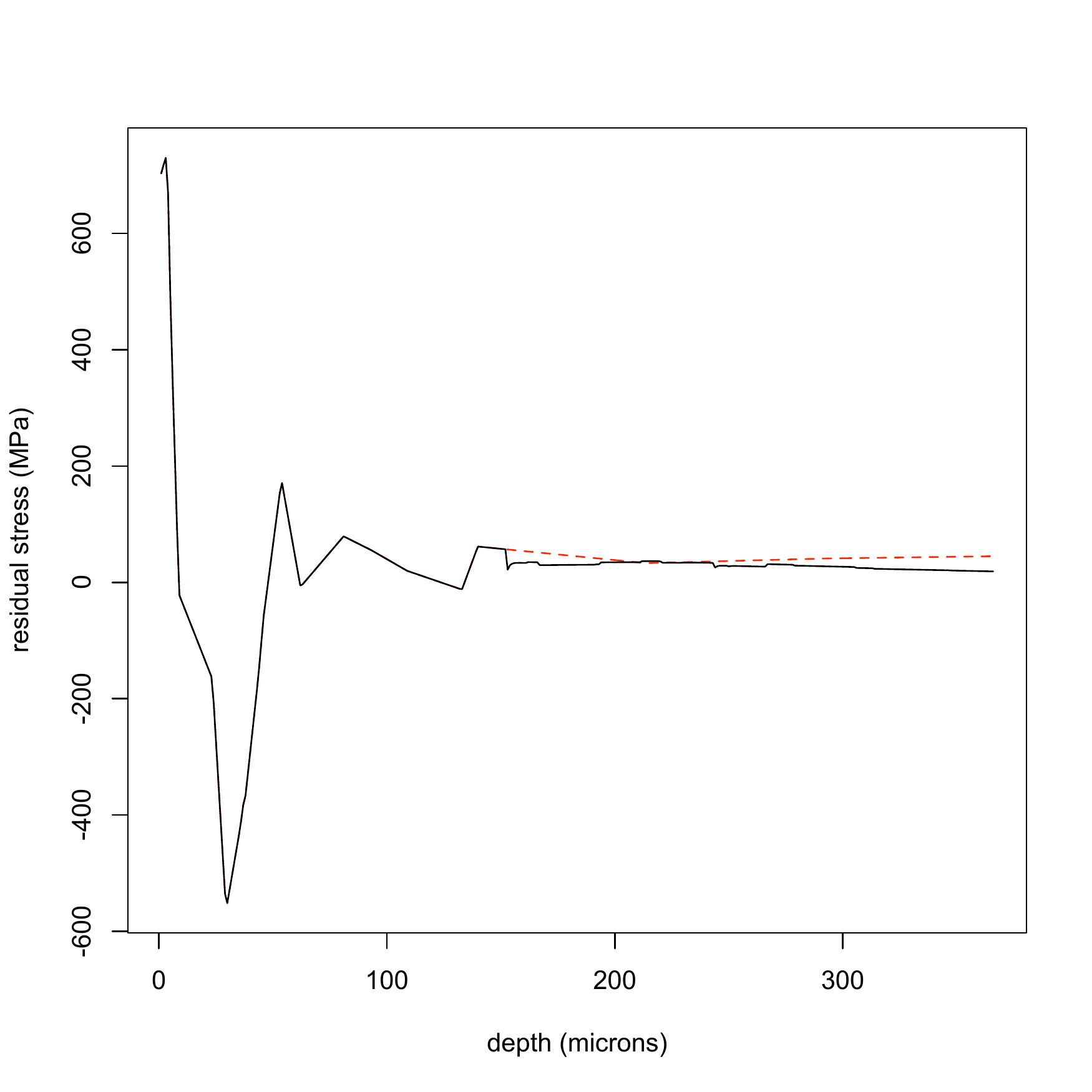}
\includegraphics[width=2in]{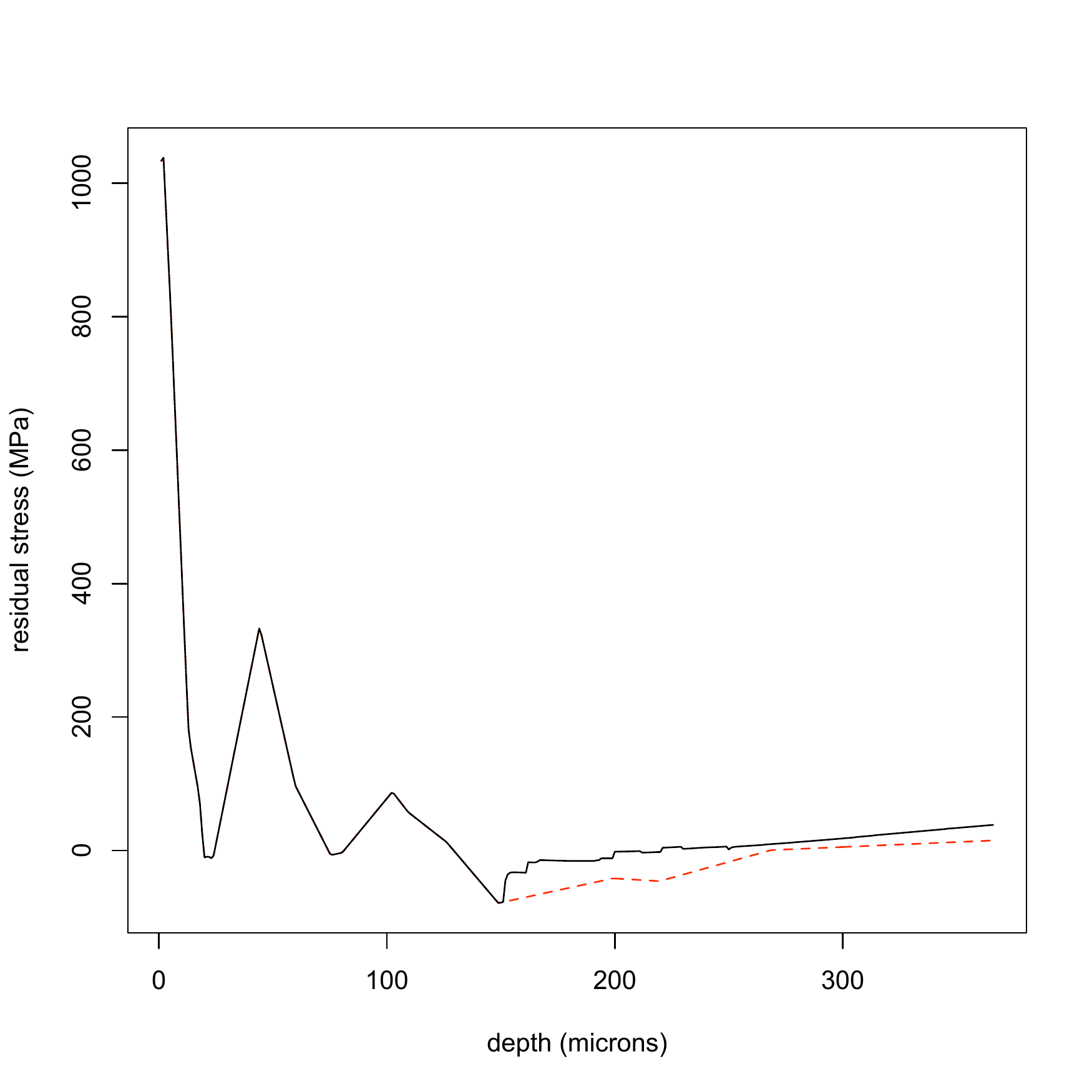}\\
\caption{Missing data estimated by EM.}\label{EMillustration}
\end{figure}

In the second stage, the proposed EM algorithm is applied. We applied 10 iterations ($q=10$) using (\ref{gibbs3}) inside each EM iteration.  With $\Delta=0.05$, the EM algorithm terminated after 21 iterations. Figure \ref{EMillustration} shows some of the missing data estimated by the procedure. They are reasonably close to the true observations (the mean squared prediction error is 45.88), showing that the procedure works well.
The maximum likelihood estimates of the correlation
parameters are given by $
\hat{\boldsymbol{\alpha}}$=$(1.34$, $1.68$, $1.47$, $3.35$, $2.80$,
$4.33$, $1.82$, $3.44$, $4.09$, $1.52$)$'$ and $\hat{\beta}=1.12$.
The computational saving obtained by the proposed EM procedure is quite substantial. The EM procedure took about 9.25 hours on a 3.4-GHz PC, whereas the naive kriging extension did not even converge after waiting for four days.

\begin{figure}[h]
\centering\resizebox{490pt}{350pt}
{\includegraphics{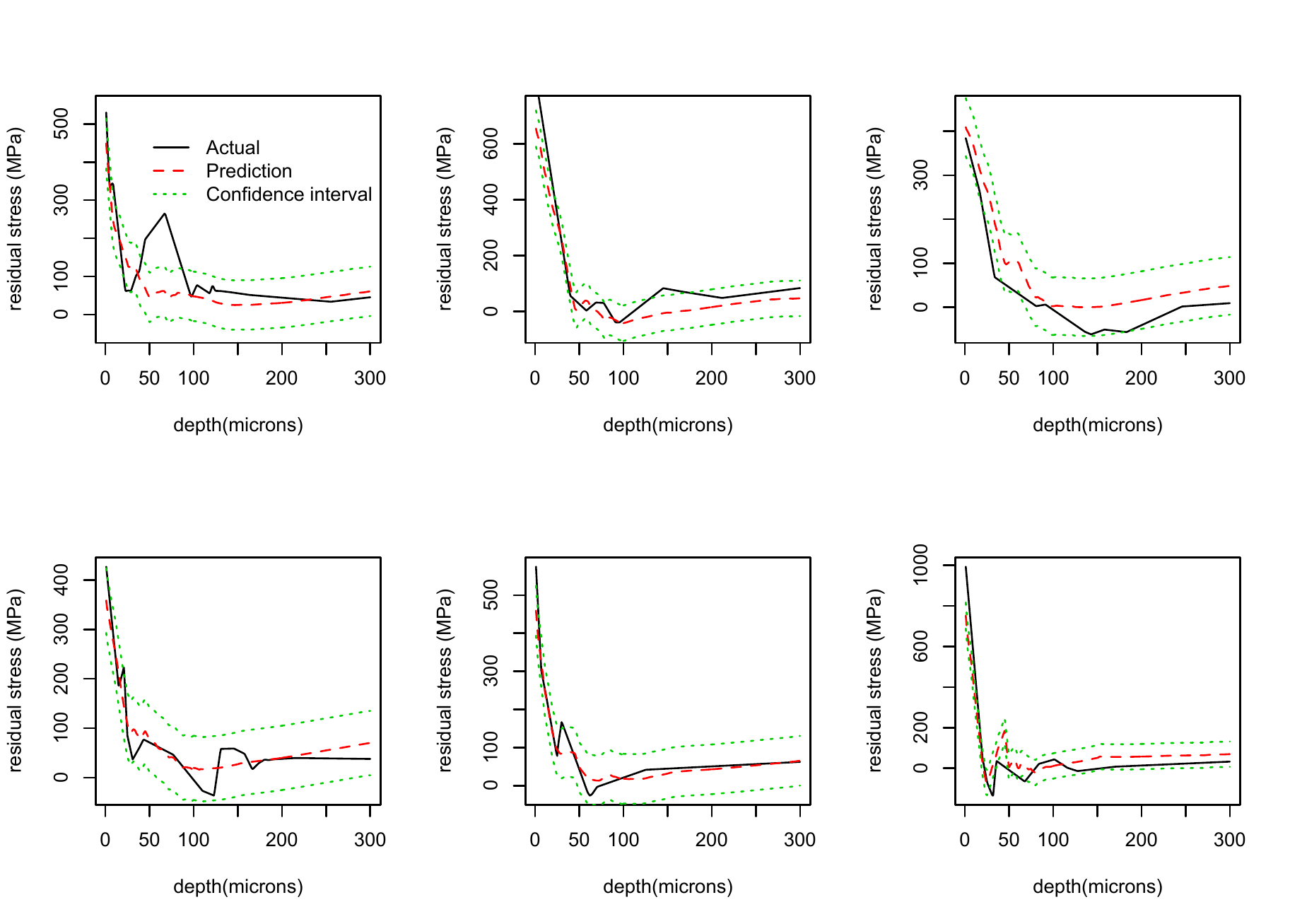}} \caption{Leave-one-out prediction and confidence interval}\label{LOONR}
\end{figure}

The $100(1-\kappa)\%$ confidence interval for the prediction  is given by (Santner et al. 2003, pp. 93),
\begin{equation}\label{CI}
\hat{y}(\boldsymbol{x},t)\pm Z_{\kappa/2}\hat{\sigma} \{ 1-\boldsymbol{r}
(\boldsymbol{x},t)' {\bf R}_{\bf{X},t}^{-1}\boldsymbol{r}
(\boldsymbol{x},t)+\boldsymbol{h}'(\boldsymbol{V}' {\bf R}_{\bf{X},t}^{-1}\boldsymbol{V})^{-1}\boldsymbol{h}\}^{1/2},
\end{equation}
where $Z_{\kappa/2}$ is the upper $\kappa/2$ critial point of the standard normal distribution and
$
\boldsymbol{h}=\boldsymbol{\upsilon}(\boldsymbol{x},t)-\boldsymbol{V}'{\bf R}_{\bf{X},t}^{-1}\boldsymbol{r}
(\boldsymbol{x},t)$.
The $95\%$ confidence intervals are illustrated using six randomly selected profiles (Figure \ref{LOONR}). Each predicted curve was evaluated by leaving the original profile out from the training set. Note that we used a small nugget term to overcome the near-singularity  in evaluating the large matrix inverse $ {\bf R}_{\bf{X},t}^{-1}$ (Gramacy and Lee 2011). The construction of the confidence intervals took about seven hours in a 2.33 GHz computer. This is not a big concern because we need to evaluate the matrix only once for constructing the confidence interval.

To illustrate the advantage of data augmentation, we compared the proposed approach with two approaches performed on the common grid (i.e., depths up to 100 microns). The two approaches include the widely used PCA (Higdon et al. 2007) and the naive kriging approach with functional argument as an additional input. In PCA, four basis functions were identified corresponding to the largest eigen values which account for more than $95\%$ of the variation in the data. To compare the performance, we revisited the six randomly selected profiles and computed the mean squared cross validation error (MSCV) averaged over the entire depth. The MSCV for the proposed kriging method is 49.11, whereas the MSCV of the PCA method is 69.51 and the naive kriging method without data augmentation is 60.23. Although the PCA can be further improved by incorporating more basis functions, the improvement is marginal (e.g. with 10 basis functions, the MSCV is reduced only to 68.94.) The comparisons based on MSCV clearly show that the proposed kriging method provides better predictions than the other two methods. This is not surprising because without data augmentation, the information from 100 to 376 microns will be completely lost leading to severe extrapolation, which can be quite inaccurate.

\begin{figure}[h]
\centering\resizebox{400pt}{300pt}
{\includegraphics{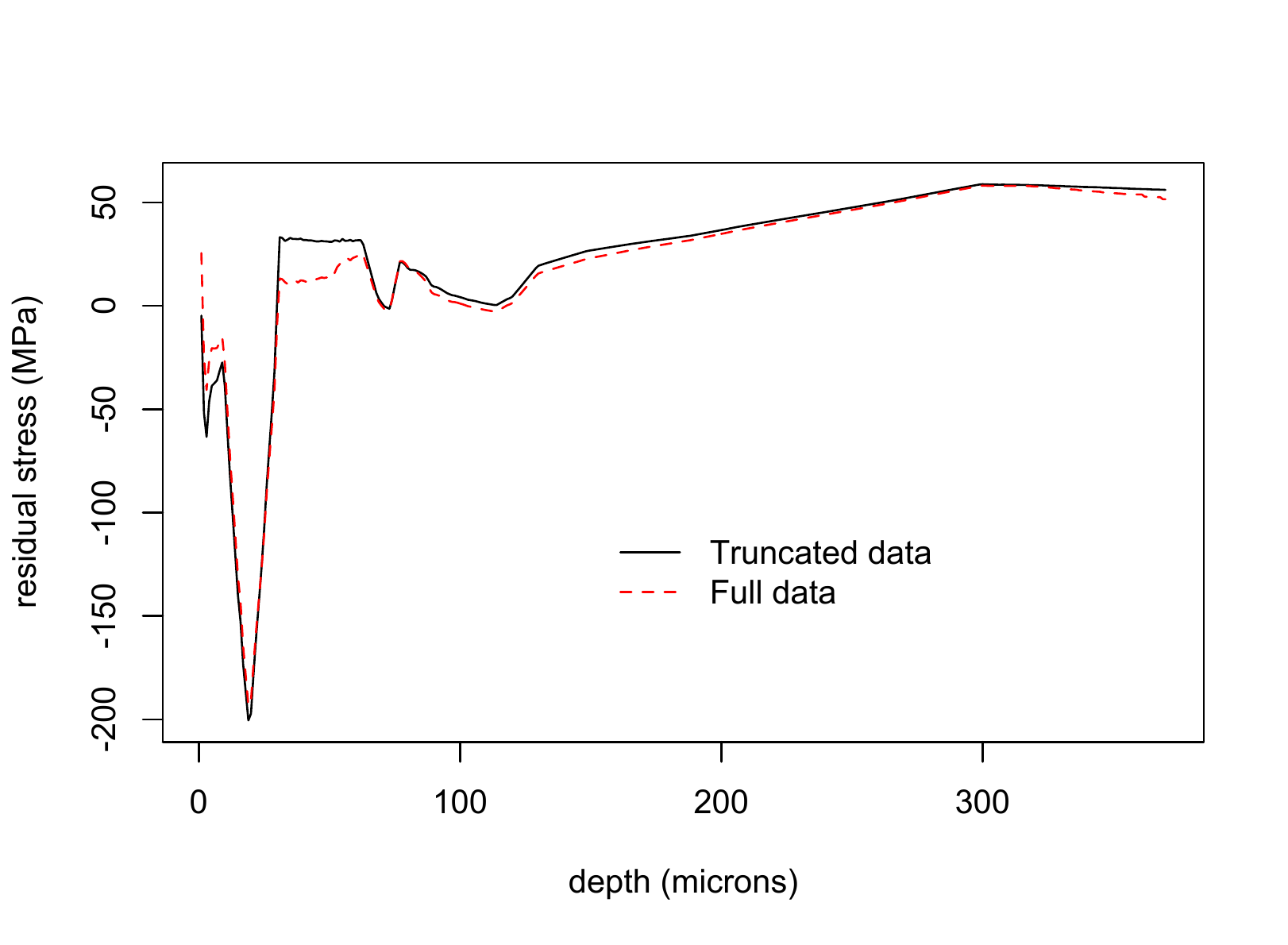}} \caption{Optimal residual stress profiles with the full regular data and the truncated irregular data.}\label{RSopt}
\end{figure}

The fitted model can be used for optimizing the residual stresses. The exact objective of optimization depends on the application area of the machined components. In most cases, the objective is to minimize the maximum tensile (i.e. maximum positive) residual stress. Thus, the objective is to find
\begin{equation}\label{minmax}
\boldsymbol{x}^*= min_{x_1,\cdots, x_9} \bigg{\{}max_{t,x_{10}} \,\,\hat{y}(\boldsymbol{x},t)\bigg{\}}.
\end{equation}
The maximization over depth ($t$) and locations ($x_{10}$) provides the largest residual stress over the machined component.
The optimal setting is given by
$\boldsymbol{x}^*$=($1.00$, $11.38$, $20.5$, $11.49,$ $9.2,$ $6.14,$ $13.11,$ $3.99,$ $9.12)'.$
Based on this setting, the predicted maximum residual stress is 58.84 MPa, which is about 53$\%$ smaller than the observed smallest maximum residual stress, 124.26 MPa, in the experiment. These results are reasonably close to that found using the original regular data, in which the optimal setting is $\boldsymbol{x}^*=(1.00$, $11.61$, $19.87$, $11.44$, $10.48$, $7.66$, $12.63$, $4.41$, $9.35)'$ and the predicted maximum residual stress is 58.15 MPa. The predicted optimal residual stress profiles are illustrated in Figure  \ref{RSopt} with the solid line indicating the prediction using the irregular data and the dash line indicating the prediction using the full regular data.

\begin{figure}[h]
\centering\resizebox{520pt}{270pt}
{\includegraphics{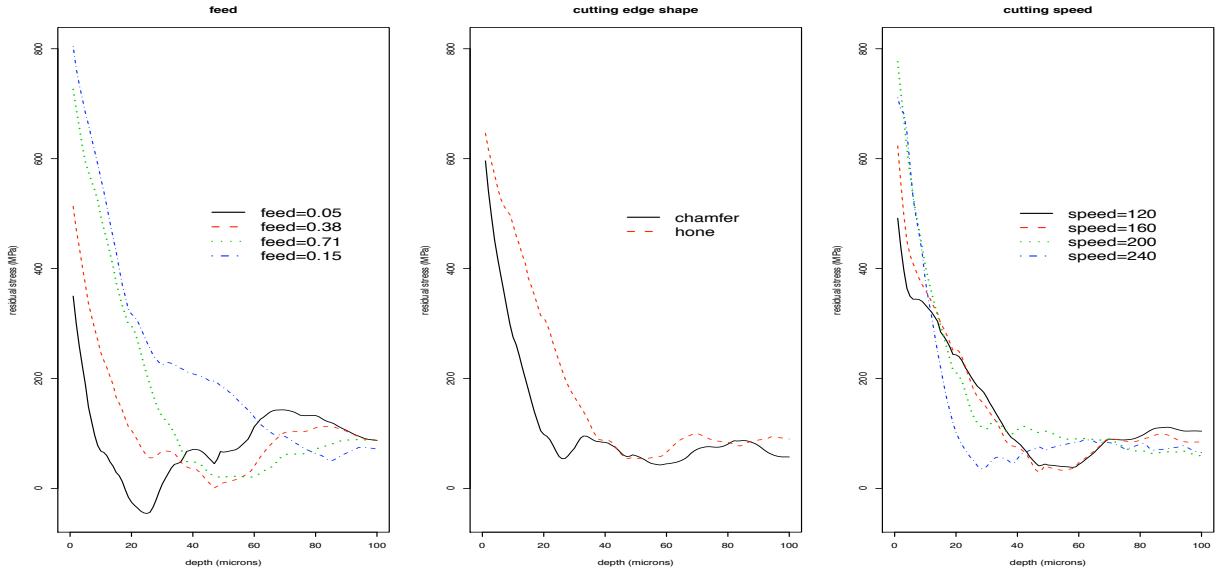}} \caption{Significant main effects on the residual stress profile.}\label{SEN}
\end{figure}

We also performed sensitivity analysis based on the main effect plots and two-factor interaction plots discussed in (Welch et al.
1992). The three main
effects of feed, cutting edge shape, and the cutting speed, appear to
be most important. Their effects on the residual stress profile
are shown in Figure \ref{SEN}. For better clarity, the residual
stress profiles were plotted with depths smaller than $100$ microns (as
the depth increases, the residual stresses converge to 0). We can see that the surface residual
stress (i.e., depth=0) increases with feed. This trend is consistent with actual observations in the process. In Figure \ref{SEN}(b), the
hone edge tool leads to larger surface residual stress than with the
chamfer edge tool. In Figure \ref{SEN}(c), the general effect of increasing the cutting speed is to cause the residual stress profile
to become more tensile (or positive). This can be physically explained on the basis of the process physics and is attributed to increased thermal effects with increase in cutting speed.

\begin{center}
{\large\bf 5. SUMMARY AND CONCLUDING REMARKS}
\end{center}

Computer experiments with functional response
are encountered in many engineering and scientific studies.
Most available methods
in the literature for analyzing computer experiments, however, focus on single outputs.
Extending these methods to functional data analysis pose severe computational challenges.
This article proposed a systematic methodology for modeling functional outputs using kriging especially for the data observed on an irregular grid. The computational challenge is successfully tackled by the proposed two-stage model building
procedure which incorporates the Kronecker product technique and a version of the EM algorithm for
estimation. A direct application of the standard EM algorithm was not possible. Therefore, we introduced  a Gibbs sampling-based computationally efficient algorithm for estimating the missing data run-by-run. The proposed method is illustrated by a
hard turning machining simulation experiment in which the functional response, residual stress, is collected over depth. Based on the
fitted model, the effects of each factor on the residual stress profile are studied  and optimal settings are identified.

In this research, the uncertainty quantification is studied based on the plug-in approach, which underestimates the true variance. A possible extension is to develop a fully Bayesian approach by incorporating prior information from the mean coefficients and correlation parameters so that the uncertainty about the unknown parameters can be taken into account. However, a direct application of the fully Bayesian approach is computationally infeasible because each MCMC sample involves a large matrix inversion. Faster Bayesian computing techniques should be considered to tackle this problem (Joseph 2012).

A weakness of the current framework is the stationarity assumption employed in the kriging model. One approach to relax this assumption is to incorporate treed partitioning based on the idea of Chipman et al. (2002) and Gramacy and Lee (2008). Although additional computational efforts are required due to the recursive partitioning of trees, extension of this procedure is computationally tractable because the proposed algorithm can be implemented within each partition (leaf of the tree), which has less data.

\begin{center}
{\bf Acknowledgments}
\end{center}

The authors thank the Editor, AE, and three referees for their valuable comments and suggestions. This research was supported by U. S. National Science Foundation grants CMMI-0654369, CMMI-0927572, and DMS-0905753, and also in part by a grant from the U. S. Army Research Laboratory and the U. S. Army Research Office under contract number W911NF-08-1-0368.

\begin{center}
{\bf Appendix A: Proof of Proposition 1}
\end{center}

The objective is to calculate conditional posterior
$f(\bm c_i|\boldsymbol{y},\bm z_{(-i)})=f(\bm c_i|\boldsymbol{y}_i,\boldsymbol{y}_{(-i)},\bm z_{(-i)})$.
We have $$f(\bm c_i|\boldsymbol{y}_i,\boldsymbol{y}_{(-i)},\bm z_{(-i)}) \propto f(\boldsymbol{y}_i,\boldsymbol{y}_{(-i)},\bm z_{(-i)}|\bm c_i) f(\bm c_i),$$
and
$$f(\boldsymbol{y}_i,\boldsymbol{y}_{(-i)},\bm z_{(-i)}|\bm c_i)=f(\boldsymbol{y}_{(-i)},\boldsymbol{z}_{(-i)}|\boldsymbol{y}_{i},\bm c_i)f(\boldsymbol{y}_{i}|\bm c_i).$$
Therefore,
\begin{equation}\label{eq1}
\begin{array}{rl}
f(\bm c_i|\boldsymbol{y}_i,\boldsymbol{y}_{(-i)},\bm z_{(-i)})&\propto f(\boldsymbol{y}_{(-i)},\bm z_{(-i)}|\boldsymbol{y}_{i},\bm c_i)f(\boldsymbol{y}_{i}|\bm c_i) f(\bm c_i)\\
& \propto {f(\boldsymbol{y}_{(-i)},\bm z_{(-i)}|\boldsymbol{y}_{i},\bm c_i)f(\bm c_i)f(\boldsymbol{y}_{i}|\bm c_i) f(\bm c_i)}/{f(\bm c_i)}.\\
\end{array}
\end{equation}
Because $\boldsymbol{y}_i \subset \bm c_i$, we have $f(\boldsymbol{y}_{(-i)},\bm z_{(-i)}|\boldsymbol{y}_{i},\bm c_i)= f(\boldsymbol{y}_{(-i)},\bm z_{(-i)}|\bm c_i)$ and equation (\ref{eq1}) can be written as
\begin{equation}\label{posteriorAppendix}
\begin{array}{rl}
f(\bm c_i|\boldsymbol{y}_i,\boldsymbol{y}_{(-i)},\bm z_{(-i)}) &\propto f(\boldsymbol{y}_{(-i)}\bm z_{(-i)}|\bm c_i)f(\bm c_i)f(\boldsymbol{y}_{i}|\bm c_i) f(\bm c_i)/{f(\bm c_i)}\\
& \propto f(\bm c_i|\boldsymbol{y}_{(-i)},\bm z_{(-i)})f(\bm c_i|\boldsymbol{y}_{i})/f(\bm c_i).
\end{array}
\end{equation}
Since the three prior distributions on the right hand side of (\ref{posteriorAppendix}) are all normally distributed as given in  (\ref{prior1}), (\ref{prior2}), and (\ref{prior3}), we have
$$
\begin{array}{rl}
f(\bm c_i|\boldsymbol{y}_i,\boldsymbol{y}_{(-i)},\bm z_{(-i)}) \propto & -\frac{1}{2} \exp\bigg{\{}({\bm c}_i-\boldsymbol{\zeta}_{(-i)})'{\bf \Sigma}_{(-i)}^{-1}({\bm c}_i-\boldsymbol{\zeta}_{(-i)})+ ({\bm c}_i-\boldsymbol{\zeta}_{i})'{\bf \Sigma}_{i}^{-1}({\bm c}_i-\boldsymbol{\zeta}_{i})\\
& - ({\bm c}_i-\boldsymbol{\zeta}_{{\bm c}_i})'{\bf \Sigma}_{\bm c_i}^{-1}({\bm c}_i-\boldsymbol{\zeta}_{{\bm c}_i})\bigg{\}}\\
\propto& -\frac{1}{2}\exp \bigg{\{} ({\bm c}_i-\boldsymbol{\eta}_{i})'\boldsymbol{\Gamma}_{i}^{-1}({\bm c}_i-\boldsymbol{\eta}_{i})\bigg{\}},\\
\end{array}
$$
where $\boldsymbol{\eta}_{i}$ and $\boldsymbol{\Gamma}_{i}$ are given in (\ref{pmean}) and (\ref{pvariance}).
Therefore, the conditional distribution of
$f(\bm c_i|\boldsymbol{y}_i,\boldsymbol{y}_{(-i)},\bm z_{(-i)})$ follows.\qed\\

\begin{center}
{\bf Appendix B: Proof of Proposition 2}
\end{center}

First, it holds that
$$
\begin{array}{rl}
&\big{(}\boldsymbol{r}_{(-i)}(\boldsymbol{x}_i,t_1),\cdots,
\boldsymbol{r}_{(-i)}(\boldsymbol{x}_i,t_m)\big{)}'\bigg{(} {\bf R}^{-1}_{\bf{X}_{(-i)}}\otimes {\bf R}^{-1}_{\bf t}\bigg{)}\\
=&\bigg{(}\boldsymbol{r}'_{(-i)}(\boldsymbol{x}_i) \otimes {\bf R}_{\bf t}\bigg{)}\bigg{(}{\bf R}^{-1}_{\bf{X}_{(-i)}} \otimes {\bf R}^{-1}_{\bf t}\bigg{)}\\
=&\bigg{(}\boldsymbol{r}'_{(-i)}(\boldsymbol{x}_i) {\bf R}^{-1}_{\bf{X}_{(-i)}}\bigg{)}\otimes \big{(}{\bf R}_{\bf t} {\bf R}^{-1}_{\bf t}\big{)}\\
=&\bigg{(}\boldsymbol{r}'_{(-i)}(\boldsymbol{x}_i) {\bf R}^{-1}_{\bf{X}_{(-i)}}\bigg{)}\otimes {\bm I}_{m \times m}.
\end{array}
$$
By taking a factor ${\bf \Sigma}_{(-i)}^{-1}$ out of (\ref{pmean}) and assuming that
$(d_1,\cdots, d_{i-1},d_{i+1},\cdots, d_n)=\boldsymbol{r}'_{(-i)}(\boldsymbol{x}_i) {\bf R}^{-1}_{\bf{X}_{(-i)}}$, we have
\begin{equation}\label{convergence1}
\begin{array}{rl}
\bm z_i^j=&\sum_{k=1}^{i-1} d_k \bigg{[}({\bm I}_{m\times m}+{\bf \Sigma}_{(-i)}{\bf \Sigma}_i^{-1}-{\bf \Sigma}_{(-i)}{\bf \Sigma}^{-1}_{\bm c_i})^{-1}\bigg{]}_{ik} \bm z_{k}^{j}\\
&+\sum_{k=i+1}^n d_k \bigg{[}({\bm I}_{m\times m}+{\bf \Sigma}_{(-i)}{\bf \Sigma}_i^{-1}-{\bf \Sigma}_{(-i)}{\bf \Sigma}^{-1}_{\bm c_i})^{-1}\bigg{]}_{ik} \bm z_{k}^{j-1}+C,
\end{array}
\end{equation}
where $C$ is independent of $\bm z_i$'s. For notational simplicity, we will denote $\big{[}\cdot\big{]}_{ik}$ by $\big{[}\cdot\big{]}$ in the remaining part of the proof.

Consider a set of $\bm z_i^j $ that satisfy (\ref{gibbs3}) simultaneously, i.e., $\bm z_i^j \rightarrow \bm z_i$, we have
\begin{equation}\label{convergence2}
\begin{array}{rl}
\bm z_i=&\sum_{k=1}^{i-1} d_k \bigg{[}({\bm I}_{m\times m}+{\bf \Sigma}_{(-i)}{\bf \Sigma}_i^{-1}-{\bf \Sigma}_{(-i)}{\bf \Sigma}^{-1}_{\bm c_i})^{-1}\bigg{]} \bm z_{k}\\
&+\sum_{k=i+1}^n d_k \bigg{[}({\bm I}_{m\times m}+{\bf \Sigma}_{(-i)}{\bf \Sigma}_i^{-1}-{\bf \Sigma}_{(-i)}{\bf \Sigma}^{-1}_{\bm c_i})^{-1}\bigg{]} \bm z_{k}+C.
\end{array}
\end{equation}
Subtracting (\ref{convergence2}) from (\ref{convergence1}) and assuming that $\boldsymbol{\varepsilon}_i^j=\bm z_i^j-\bm z_i$, we obtain
$$
\begin{array}{rl}
\boldsymbol{\varepsilon}_i^j=&\sum_{k=1}^{i-1} d_k \bigg{[}({\bm I}_{m\times m}+{\bf \Sigma}_{(-i)}{\bf \Sigma}_i^{-1}-{\bf \Sigma}_{(-i)}{\bf \Sigma}^{-1}_{\bm c_i})^{-1}\bigg{]} \boldsymbol{\varepsilon}_{k}^j\\
&+\sum_{k=i+1}^n d_k \bigg{[}({\bm I}_{m\times m}+{\bf\Sigma}_{(-i)}{\bf \Sigma}_i^{-1}-{\bf \Sigma}_{(-i)}{\bf \Sigma}^{-1}_{\bm c_i})^{-1}\bigg{]} \boldsymbol{\varepsilon}_{k}^{j-1}.
\end{array}
$$
Define $\delta=\mbox{max}_i \sum_{k=1,k\neq i}^n d_k \bigg{|}\bigg{|} \big{[}{\bm I}_{m\times m}+{\bf \Sigma}_{(-i)}{\bf \Sigma}_i^{-1}-{\bf \Sigma}_{(-i)}{\bf \Sigma}^{-1}_{\bm c_i}\big{]}^{-1} \bigg{|} \bigg{|}_2$.
Now we show  by induction on $i$ that the following result holds:
\begin{equation}\label{convergence3}
\mbox{max}_i \parallel\boldsymbol{\varepsilon}_i^j\parallel \leq \delta \,\,\mbox{max}_i \parallel\boldsymbol{\varepsilon}_i^{j-1}\parallel, \,\,\, j=1,2,\cdots.
\end{equation}
For $i=1$,
$$
\boldsymbol{\varepsilon}_1^j=\sum_{k=i+1}^n d_k \big{[}{\bm I}_{m\times m}+{\bf \Sigma}_{(-i)}{\bf \Sigma}_i^{-1}-{\bf \Sigma}_{(-i)}{\bf \Sigma}^{-1}_{\bm c_i}\big{]}^{-1} \boldsymbol{\varepsilon}_{k}^{j-1},
$$
and thus, (\ref{convergence3}) clearly holds. Assume that (\ref{convergence3}) holds for $i=1,\cdots, l-1$ and based on assumption (\ref{prop2assump}), we have for $i=l$
$$
\begin{array}{rl}
\parallel\boldsymbol{\varepsilon}_l^j\parallel\leq&\delta\,\,\mbox{max}_l \parallel\boldsymbol{\varepsilon}_l^{j-1}\parallel\sum_{k=1}^{l-1} d_k \bigg{|}\bigg{|}\big{[}({\bm I}_{m\times m}+{\bf\Sigma}_{(-l)}{\bf \Sigma}_l^{-1}-{\bf \Sigma}_{(-l)}{\bf \Sigma}^{-1}_{\bm c_l}\big{]}^{-1} \bigg{|}\bigg{|}_2\\
&+\sum_{k=l+1}^n d_k \bigg{|}\bigg{|}\big{[}{\bm I}_{m\times m}+{\bf \Sigma}_{(-i)}{\bf \Sigma}_l^{-1}-{\bf \Sigma}_{(-l)}{\bf \Sigma}^{-1}_{\bm c_l}\big{]}^{-1}\bigg{|}\bigg{|}_2 \parallel\boldsymbol{\varepsilon}_{k}^{j-1}\parallel\\
< &\delta \,\,\mbox{max}_l \parallel\boldsymbol{\varepsilon}_l^{j-1}\parallel\sum_{k=1,k\neq l}^{n} d_k \bigg{|}\bigg{|}\big{[}({\bm I}_{m\times m}+{\bf \Sigma}_{(-l)}{\bf \Sigma}_i^{-1}-{\bf \Sigma}_{(-l)}{\bf \Sigma}^{-1}_{\bm c_l}\big{]}^{-1}\bigg{|}\bigg{|}_2 \\
\leq & \delta\,\, \mbox{max}_l \parallel\boldsymbol{\varepsilon}_l^{j-1}\parallel.
\end{array}
$$
The result in (\ref{convergence3}) implies that $\mbox{max}_i \parallel\boldsymbol{\varepsilon}_i^j\parallel \leq \delta^j \,\,\mbox{max}_i \parallel\boldsymbol{\varepsilon}_i^{0}\parallel$ and $\mbox{max}_i \parallel\boldsymbol{\varepsilon}_i^j\parallel \rightarrow 0$ as $j\rightarrow \infty$.
Therefore, $\bm z_i^j$'s converge to the values that satisfy equations (\ref{gibbs3}) simultaneously.
According to the result in Yee et al. (2002), $\bm z_i^j$ converge to the posterior mean $E(\bm z_i|\boldsymbol{y},\hat{\boldsymbol{\theta}}^{(k)})$ when $m \rightarrow \infty$.\qed

\begin{center}
{\large\bf References}
\end{center}
\begin{description}

\item An, J. and Owen, A. (2001), ``Quasi-Regression," \emph{Journal of Complexity}, 17, 588-607.

\item Banerjee, S., Gelfand, A. E., Finley, A. O., and Sang, H. (2008), ``Gaussian Predictive Process Models for Large Spatial Datasets,"
\emph{Journal of the Royal Statistical Society, Series B}, 70, 825-848.

\item Bayarri, M. J., Berger, J. O., Cafeo, J., Garcia-Donato, G.,
Liu, F., Palomo, Parthasarathy, R.J., Paulo, R., Sacks, J., Walsh,
D. (2007), ``Computer Model Validation with Functional Output,"
\emph{Annals of Statistics}, 35, 1874-1906.

\item Bayarri, M. J., Berger, J. O., Kennedy, M. C., Kottas, A., Paulo, R., Sack, J., Cafeo, J. A., Lin C.-H., and Tu, J. (2009),
``Predicting Vehicle Crashworthiness: Validation of Computer Models for Functional and Hierarchical Data,"
\emph{Journal of the American Statistical Association}, 104, 929-943.


\item Chan, K. S. and Ledolter, J. (1995), ``Monte Carlo EM Estimation for Time Series Models Involving Counts," \emph{Journal of the American Statistical Association}, 90, 242-252.

\item Chipman, H., George, E., and McCulloch, R. (2002), ``Bayesian Treed models," \emph{Machine Learning}, 48, 303-324.

\item Conti, S., Gosling, J. P., Oakley, J. E., and O'Hagan, A. (2009), ``Gaussian Process Emulation of Dynamic Computer Codes," \emph{Biometrika}, 96, 663-676.

\item Conti, S. and O'Hagan, A. (2010), ``Bayesian Emulation of Complex Multi-Output and Dynamic Computer Models," \emph{Journal of Statistical Planning and Inferenec}, 140, 640-651.

\item Cressie, N. A. (1993), \emph{Statistics for Spatial Data}, New York: John Wiley.

\item Currin, C., Mitchell, T. J., Morris, M. D., and Ylvisaker, D. (1991), ``Bayesian Prediction of Deterministic Functions with Applications to the Design and Analysis of Computer Experiments," \emph{Journal of the American Statistical Association}, 86, 953-963.

\item Dempster, A., Laird, N., and Rubin, D. (1977), ``Maximum Likelihood from Incomplete Data via the EM Algorithm
(with Discussion)," \emph{Journal of the Royal Statistical Society:} Series B, 39, 1-38.

\item Efron, B. (1994),``Missing Data, Imputation, and the Bootstrap," \emph{Journal of the American
Statistical Association}, 89, 463-475.

\item Fang, K. T., Li, R., and Sudjianto, A. (2006), \emph{Design
and Modeling for Computer Experiments}, New York: CRC Press.

\item Finley, A. O., Banerjee, S., Waldmann, P. and Ericsonn, T. (2009), ``Hierarchical Spatial Modeling of Additive and
Dominance Genetic Variance for Large Spatial Trial Datasets," \emph{Biometrics}, 61, 441–451.

\item Fort, G. and Moulines, E. (2003), ``Convergence of the Monte Carlo
Expectation Maximization for Curved Exponential Families," \emph{Annals of Statistics}, 31, 1220-1259.

\item Fuentes, M. (2007), ``Approximate Likelihood for Large Irregularly Spaced Spatial Data," \emph{Journal of the American Statistical Association}, 102, 321-331.


\item Gramacy, R. B. and Lee, H. K. H. (2008), ``Bayesian Treed Gaussian Process Models with an Application to Computer Modeling," \emph{Journal of the American Statistical Association}, 103, 11191130.

\item Gramacy, R. B. and Lee, H. K. H. (2012), ``Cases for the nugget in modeling computer experiments," \emph{Statistics and Computing}, to appear.


\item Higdon, D., Gattiker, J., Williams, B., and Rightley, M. (2007),
``Computer Model Validation Using High Dimensional Outputs,"
\emph{Bayesian Statistics} 8, eds. Bernardo, J., Bayarri, M. J.,
Dawid, A. P., Berger, J. O., Heckerman, D., Smith, A. F. M., and
West, M., London: Oxford University Press.

\item Hung, Y., Joseph, V. R., and Melkote, S. N. (2009), ``Design and Analysis of Computer Experiments with Branching and Nested Factors," \emph{Technometrics}, 51, 354-365.

\item Joseph, V. R. (2012), ``Bayesian Computation Using Design of Experiments-based Interpolation Technique," \emph{Technometrics}, to appear.


\item Joseph, V. R., Hung, Y., and Sudjianto, A. (2008), ``Blind Kriging: A New Method for Developing Metamodels,"
\emph{ASME Journal of Mechanical Design}, 130, 031102-1-8.

\item Joseph, V. R. and Kang, L. (2011), ``Regression-Based Inverse Distance Weighting with Applications to Computer Experiments,'' \emph{Technometrics}, 53, 254-265.

\item Kaufman, C., Schervish, M., and Nychka, D. (2008) ``Covariance Tapering for Likelihood-Based Estimation in Large Spatial Datasets." \emph{Journal of the American Statistical Association}, 103, 15561569.

\item Kennedy, M. C. and O'Hagan, A. (2001), ``Bayesian Calibration of Computer Models," \emph{Journal of the Royal Statistical Society: Series B}, 63, 425-464.

\item Liu, F., Bayarri, M.J., Berger, J., Paulo, R. and Sacks, J.
(2008), ``A Bayesian Analysis of the Thermal Challenge Problem,"
\emph{Computer Methods in Applied Mechanics and Engineering}, 197,
2457-2466.

\item Liu, F. and West, M. (2009), ``A Dynamic Modeling Strategy for Bayesian Computer Model Emulation," \emph{Bayesian Analysis}, 4, 393-412.

\item Liu, X. Chen, W ., and Tsui, K.-L., (2008), ``Regression Modeling for Computer Model Validation with Functional Responses,"  \emph{Proceedings of the ASME 2008 International Design Engineering Technical Conferences and Computers and Information in Engineering Conference}, Paper No. DETC2008-49662, Brooklyn , New York.

\item Miller, S. and Shih, A. (2007),``Thermo-Mechanical Finite Element Modeling of the Friction Drilling Process," \emph{ASME Journal of Manufacturing Science and Engineering}, 129, 531-538.




\item Puppu, V. and Bagchi, P. (2008),``3D Computational Modeling and Simulation of Leukocyte Rolling Adhesion and Deformation," \emph{Computers in Biology and Medicine}, 38, 738-754.

\item Ramsay, J. O. and Silverman, B. W. (2005), \emph{Functional Data Analysis}, 2nd ed. New York: Springer.

\item Rougier, J. (2008), ``Efficient Emulators for Multivariate Deterministic Functions," \emph{Journal
of Computational and Graphical Statistics}, 17, 827-843.

\item Sacks, J., Welch, W. J., Mitchell, T. J. and Wynn, H. P.
(1989), ``Design and Analysis of Computer Experiments,"
\emph{Statistical Science}, 4, 409-423.

\item Santner, T. J., Williams, B. J., and Notz, W. I. (2003),
\emph{The Design and Analysis of Computer Experiments},
New York: Springer.


\item Williams, B., Higdon, D., Gattiker, J., Moore, L., McKay, M., and Keller-MacNulty, S. (2006),
``Combining Experimental Data and Computer Simulations, With and Application to Flyer Plate Experiments,"
\emph{Bayesian Analysis}, 1, 765-792.

\item Wu, C. F. J. (1983), ``On the Convergence Properties of the EM Algorithm," \emph{Annals of Statistics}, 11, 95-103.

\item Yee, L., Johnson, W. O., and Samaniego, F. J. (2002), `` Asymptotic Approximations to Posterior Distributions via Conditional Moment
Equations," \emph{Biometrika}, 84, 755-767.
\end{description}

\end{document}